\documentclass{ws-ijns}
\usepackage{graphicx}
\usepackage[english]{babel}
\usepackage[super,sort,compress]{cite}
\usepackage{amsmath}
\usepackage{mathtools}

\usepackage{multirow}
\usepackage[flushleft]{threeparttable}
\usepackage[normalem]{ulem}

\usepackage{enumitem}
\newlist{tabitem}{itemize}{1}
\setlist[tabitem]{wide=0pt, nosep, leftmargin= * ,label=\textbullet,after=\vspace{-\baselineskip},before=\vspace{-2\baselineskip}}

\usepackage{multirow}
\usepackage{tabularx,colortbl}
\def\cpar{\hss\egroup\line\bgroup\hss}

\usepackage{hhline}

\usepackage[T1]{fontenc}
\usepackage{hyperref}

\begin{document}

%\catchline{0}{0}{2005}{}{}

\markboth{Peh et al.}{Seizure Detection}

\title{Six-center Assessment of CNN-Transformer with Belief Matching Loss for Patient-independent Seizure Detection in EEG}

\author{Wei Yan Peh$^{1,2}$, 
Prasanth Thangavel$^{1,2}$, 
Yuanyuan Yao$^{3}$,
John Thomas$^{4}$, 
Yee-Leng Tan$^{5}$,
and Justin Dauwels$^{6*}$}

\footnotetext{Corresponding author, Email: j.h.g.dauwels@tudelft.nl}
\footnotetext{An extended version of this paper can be found in \href{https://arxiv.org/abs/2208.00025}{https://arxiv.org/abs/2208.00025}.}

\address{\vspace{0.5 cm}
$^{1}$Nanyang Technological University (NTU), Interdisciplinary Graduate School (IGS), Singapore 639798\\
$^{2}$NTU, School of Electrical and Electronic Engineering (EEE), Singapore 639798\\
$^{3}$Katholieke Universiteit Leuven, Oude Markt 13, 3000 Leuven, Belgium\\
$^{4}$Montreal Neurological Institute (MNI), McGill University, Montreal, QC H3A 2B4, Canada\\
$^{5}$National Neuroscience Institute (NNI), Singapore 308433\\
$^{6}$Delft University of Technology (TU Delft), Department of Microelectronics, 2628 CD Delft, Netherlands}

\maketitle

\begin{abstract}
Neurologists typically identify epileptic seizures from electroencephalograms (EEGs) by visual inspection. This process is often time-consuming, especially for EEG recordings that last hours or days. To expedite the process, a reliable, automated, and patient-independent seizure detector is essential. However, developing a patient-independent seizure detector is challenging as seizures exhibit diverse characteristics across patients and recording devices. In this study, we propose a patient-independent seizure detector to automatically detect seizures in both scalp EEG and intracranial EEG (iEEG). First, we deploy a convolutional neural network with transformers and belief matching loss to detect seizures in single-channel EEG segments. Next, we extract regional features from the channel-level outputs to detect seizures in multi-channel EEG segments. At last, we apply postprocessing filters to the segment-level outputs to determine seizures' start and end points in multi-channel EEGs. Finally, we introduce the minimum overlap evaluation scoring as an evaluation metric that accounts for minimum overlap between the detection and seizure, improving upon existing assessment metrics. We trained the seizure detector on the Temple University Hospital Seizure (TUH-SZ) dataset and evaluated it on five independent EEG datasets. We evaluate the systems with the following metrics: sensitivity (SEN), precision (PRE), and average and median false positive rate per hour (aFPR/h and mFPR/h). Across four adult scalp EEG and iEEG datasets, we obtained SEN of 0.617-1.00, PRE of 0.534-1.00, aFPR/h of 0.425-2.002, and mFPR/h of 0-1.003. The proposed seizure detector can detect seizures in adult EEGs and takes less than 15s for a 30 minutes EEG. Hence, this system could aid clinicians in reliably identifying seizures expeditiously, allocating more time for devising proper treatment. 
\end{abstract}


\begin{multicols}{2}

\section{Introduction}
{E}{pilepsy} is a brain disorder characterized by the manifestations of sudden unprovoked seizures~\cite{jirsa2014nature}. Seizures are diverse and vary significantly across patients in etiology, severity, and symptoms~\cite{nunes2012diagnosis}. Most electrographic seizures last from 30 seconds to two minutes, where a seizure lasting longer than five minutes is a medical emergency~\cite{jenssen2006long}. Epilepsy is diagnosed when a patient experiences two or more recurring seizures~\cite{goldenberg2010overview}. Around 1\% of the world population is diagnosed with epilepsy~\cite{world2005atlas}. Moreover, approximately 10\% of the population will experience a seizure within their lifetime~\cite{ferri2019ferri}. Overall, provoked and unprovoked seizures occur in about 3.5 and 4.2 per 10000 individuals annually, respectively~\cite{world2005atlas}. After a seizure episode, the likelihood of encountering another seizure event increases to about 50\%, bringing the individual to a much greater risk of relapsing~\cite{berg2008risk}.

To detect seizures, an electroencephalogram (EEG) can be utilized to measure the electrical activity in the brain~\cite{world2005atlas}. Scalp EEG records the brain activity with surface electrodes, while intracranial EEG (iEEG) measures the signals directly via implanted electrodes~\cite{mormann2007seizure}. However, visual inspection of EEGs can be time-consuming~\cite{geut2017detecting}. There is a need for automated detectors that can detect seizures reliably and quickly. Most progress has been made toward patient-specific detectors, as seizure morphologies vary across patients. Consequently, designing a seizure detector that can detect seizures in any patient can be challenging but tremendously helpful for clinicians. 

In recent studies on automated seizure detection from EEG, the detectors are validated mainly on two public seizure datasets: the Temple University Hospital seizure (TUH-SZ) dataset~\cite{shah2017optimizing, ayodele2020supervised, roy2021evaluation} and the Children’s Hospital Boston Massachusetts Institute of Technology (CHB-MIT) dataset~\cite{furbass2015prospective, mansouri2019online, gomez2020automatic, ayodele2020supervised}. In many studies, different models are proposed, including {wavelet analysis~\cite{faust2015wavelet, adeli2003analysis, ghosh2007mixed}}, {machine learning models~\cite{savadkoohi2020machine}}, {convolutional neural networks (CNNs)~\cite{furbass2015prospective, shah2017optimizing, gomez2020automatic, roy2021evaluation, ansari2019neonatal}}, {recurrent neural networks (RNNs)~\cite{shah2017optimizing}}, {long short-term memory (LSTM)~\cite{hu2020scalp}}, {transformer~\cite{bhattacharya2021epileptic}}, {transfer learning~\cite{raghu2020eeg, saab2020weak, emami2019seizure, nogay2020detection}}, quickest detection~\cite{santaniello2011quickest}, and {temporal graph convolutional networks (TGCNs)~\cite{covert2019temporal}}. 

The seizure detectors proposed in these studies are similar in architecture and/or implementation. The detectors first divide the EEGs into short multi-channel segments (segment-level), before classifying each segment as normal against seizure. Then, using the segment-level outputs, they determine the start and end points of the seizures in full EEGs. The main innovation in these studies lies in the design of the segment-level detector, where most studies propose increasingly deep and complex neural networks with millions of parameters~\cite{bhattacharya2021epileptic, covert2019temporal}.

Unfortunately, computationally intensive models may not necessarily improve patient-independent seizure detection due to the associated increased risk of overfitting~\cite{roy2019chrononet, covert2019temporal}. To resolve the bottleneck, we require a fresh perspective on this problem. As we will explain in the following, we address certain drawbacks of existing seizure detectors and resolve some of their weaknesses in this study.

First, most modern seizure detectors identify seizures at the segment-level directly. Since these detectors are trained on multi-channel EEG segments, they can only handle a fixed number of EEG electrodes (e.g., 21). To apply those models to EEGs with a different number of electrodes (e.g., 32), the models need to be retrained. In practice, the number of electrodes may vary, and this limitation is a severe impediment to clinical applications. 

To overcome this, we proposed a seizure detector that starts by detecting seizures in single-channel segments (channel-level detection). We evaluate three variations of CNN for the channel-level detector: CNN with softmax loss (CNN-SM), CNN with belief matching (BM) loss (CNN-BM), and a CNN cascaded with a transformer and BM loss (CNN-TRF-BM). The BM loss is used to improve confidence performance, making the distribution of the probability predicted similar to the actual distribution of probability observed in training data. Meanwhile, the transformer is deployed to extract long-range patterns across the signals via self-attention, which the CNNs cannot. Several existing studies have proposed detectors that detect seizures at the channel-level~\cite{lu2018classification, acharya2018deep, liu2022patient}. However, some of these also analyzed single-channel EEGs instead of multi-channel EEGs~\cite{lu2018classification}. Consequently, there is no segment-level detection, making them unsuitable for detecting seizures in multi-channels EEGs.

To resolve the restriction on the fixed number of channels, we aggregate the channel-level outputs and group them into five distinct brain regions. Then, we compute statistical features from each region, which can be done for an arbitrary number of electrodes. This approach allows us to apply the detectors to EEGs with any number of electrodes and both scalp EEG and iEEGs. In this study, we trained the proposed seizure detector on a large scalp EEG dataset (TUH-SZ dataset) and evaluated it on five independent scalp EEG and iEEG datasets. In comparison, existing seizure detectors for scalp EEGs and iEEGs are often trained and analyzed separately~\cite{bhattacharya2021epileptic}.

Finally, a good evaluation metric to measure the effectiveness of seizure detectors is necessary. Such metrics score a detection from the automated system based on how much it overlaps with a manually annotated seizure(s). Unfortunately, most studies use different evaluation approaches to assess the detectors, making comparison studies challenging. Several evaluation metrics have been proposed, including epoch-based sampling (EBS), any-overlap (OVLP), time-aligned event scoring (TAES)~\cite{shah2021objective}, and increased margin scoring (IMS)~\cite{reus2022automated}. However, these metrics do not reflect real-world clinical requirements.

For instance, Reus~\textit{et al.}~and Koren~\textit{et al.}~only reported IMS, which consider a detection correct as long as the detection is within 30s and 120s before the start or after the end of the seizure, respectively. Allowing this significant error margin could lead to huge uncertainty and low precision during detection. Meanwhile, Fürbass~\cite{furbass2015prospective} determine that a seizure is detected as long as a detection appears within a seizure event. These approaches ignored the amount of overlap required, making their measurement approach extremely lenient. Either way, it is inappropriate in clinical practice.

Therefore, we introduce the minimum overlap evaluation scoring (MOES), which requires the detection from the automated system to have a minimum overlap duration of 10s and a minimum overlap of 30\% with a ground truth seizure for it to be considered correct. In contrast, OVLP and TAES require a non-zero (e.g., 0.1\%) and perfect (100\%) overlap, respectively, which tends to under- or over-penalize the detector. By requiring a non-trivial overlap, albeit not necessarily a perfect overlap, the MOES metric has an adequate tolerance for clinical practice.

In summary, this paper performs the following:

\begin{enumerate}[leftmargin=*]

    \item We developed a patient-independent seizure detector that can be applied to scalp EEG and iEEG, regardless of the number of electrodes.

    \item We utilize a BM loss to improve the calibration performance, which is critical for decision-making. However, such approaches are rarely applied in EEG analysis, as most studies favour softmax (SM) loss. Unfortunately, many existing classification algorithms are not optimized for obtaining accurate probabilities, and their predictions may be miscalibrated.

    \item We apply CNN with transformers as a transformer can extract long-range patterns, which a CNN cannot. Transformers had been explored for seizure detection (see~\cite{bhattacharya2021epileptic}) but have yet to apply at the channel-level.

    \item We train the proposed detector on one scalp EEG dataset and test it on five independent scalp EEG and iEEG datasets. Seizure detectors are usually not assessed simultaneously on multiple datasets and not on scalp EEGs and iEEGs.

    \item We introduce the minimum overlap evaluation scoring (MOES) to assess the performance of seizure detectors. In contrast to existing metrics, the MOES metric requires a non-trivial but not necessarily perfect overlap between the detection and ground truth seizure(s) for the detection to be considered correct. Existing metrics are too lenient or strict on the overlap criteria, resulting in inaccurate results.

\end{enumerate}

\section{Materials and Methods}

\subsection{Dataset}
We analyze six public EEG datasets in this study:

\begin{enumerate}[leftmargin=*]

  \item Temple University Hospital Seizure (TUH-SZ) dataset~\cite{shah2018temple}

  \item Children’s Hospital Boston Massachusetts Institute of Technology (CHB-MIT) dataset~\cite{shoeb2004patient}

  \item Helsinki University Hospital (HUH) dataset~\cite{stevenson2019dataset}

  \item Sleep Wake Epilepsy Center at ETH Zurich (SWEC-ETHZ) dataset~\cite{burrello2019hyperdimensional}

  \item International Epilepsy Electrophysiology Portal (IEEGP) dataset~\cite{wagenaar2013multimodal}

  \item Epilepsy iEEG Multicenter (EIM) dataset~\cite{li2021neural}

\end{enumerate}

\begin{table*}
\centering
\tbl{Information on the six scalp EEG and iEEG datasets analyzed in the study. 
\label{tab:dataset_seizure_comparison}}
{
\scalebox{0.65}{

\begin{tabular}{cccccccc} 
\hline \hline
\textbf{Information} & \textbf{Details} & \textbf{TUH-SZ} & \textbf{CHB-MIT} & \textbf{\textbf{HUH}} & \textbf{SWEC-ETHZ} & \textbf{IEEGP} & \textbf{EIM} \\ 
\hline \hline

\multirow{8}{*}{\begin{tabular}[c]{@{}c@{}}EEG\\Details\end{tabular}} & Patient Type & Human & Human & Human & Human & Human/Dog & Human \\
 & Patient Age Group & Adult & Paediatric & Neonatal & Adult & Adult & Adult \\
 & EEG Type & scalp EEG & scalp EEG & scalp EEG & iEEG & iEEG & iEEG \\
 & $F_s$ (Hz) & 250-1000 & 256 & 256 & 512 & 400-5000 & 250-1000 \\
 & Channel Name & Available & Available & Available & Unavailable & Unavailable & Unavailable \\
 & Channel-level Annotation & Yes & No & No & No & No & No \\
 & Seizure Label, Type & Yes, 8 & No & No & No & No & No \\
 & No of Channels & 19,21 & 23,24,26 & 21 & 36-100 & 16-72 & 53-216 \\ 
\hline

\multirow{5}{*}{\begin{tabular}[c]{@{}c@{}}Number of \\Patients and EEGs\end{tabular}} & Patients & 637 & 24 & 75 & 16 & 12 & 31 \\
 & All EEGs & 5,610 & 683 & 75 & 100 & 12 & 102 \\
 & Non-Seizure EEGs & 4,450 & 545 & 22 & 0 & 0 & 0 \\
 & Seizure EEGs & 1,150 & 138 & 54 & 100 & 12 & 102 \\
 & Seizure Events & 3,050 & 185 & 517 & 100 & 12 & 102 \\ 
\hline

\multirow{5}{*}{Duration} & All EEGs (in hours) & 922 & 980 & 114 & 13.5 & 7.20 & 7.96 \\
 & Non-SZ EEGs (in hours) & 681 & 792 & 35.0 & 0 & 0 & 0 \\
 & SZ EEGs (in hours) & 242 & 188 & 78.6 & 13.5 & 7.20 & 7.96 \\
 & Average (All) (in minutes) & 9.84 & 86.1 & 89.64 & 8.1 & 36 & 4.68 \\
 & Average SZ (in seconds) & 54.3 & 54.4 & 90.5 & 95.9 & 37.3 & 103.7 \\
\hline \hline

\end{tabular}
}}
\end{table*}

Information about the six datasets is summarized in Table~\ref{tab:dataset_seizure_comparison}. The TUH-SZ dataset is the largest among those six datasets, with the most annotated seizure events. Hence, we utilized the TUH-SZ dataset as the primary source to train the entire seizure detector pipeline. 

Firstly, the seizure detector is trained and evaluated with the TUH-SZ dataset via 4-fold cross-validation (CV). We assign approximately the same number of patients and seizures to each fold. Next, using the trained detector, we further assess it on five other independent EEG datasets. In this way, we examine the generalizability of the detector on different EEG datasets with different EEG types and patient age groups.

For all the EEGs, a $4^{th}$ order Butterworth notch filter at 60Hz (USA) and 50Hz (EU) is applied to remove electrical interference~\cite{thomas2021automated}. Next, a 1Hz high-pass filter ($4^{th}$ order) is implemented to reject DC shifts and baseline fluctuations~\cite{thangavel2021time}. Finally, all the EEGs are downsampled to a sampling frequency $\text{F}_{s}$ of 128Hz. At last, we convert all scalp EEGs to bipolar montage, as the TUH-SZ dataset is annotated in the bipolar montage. As the montage for the iEEGs is incompatible with the bipolar montage, we keep the montage of the iEEGs at monopolar.

\begin{figurehere}
\centering
\vspace{0.5cm}
\includegraphics[width=6cm]{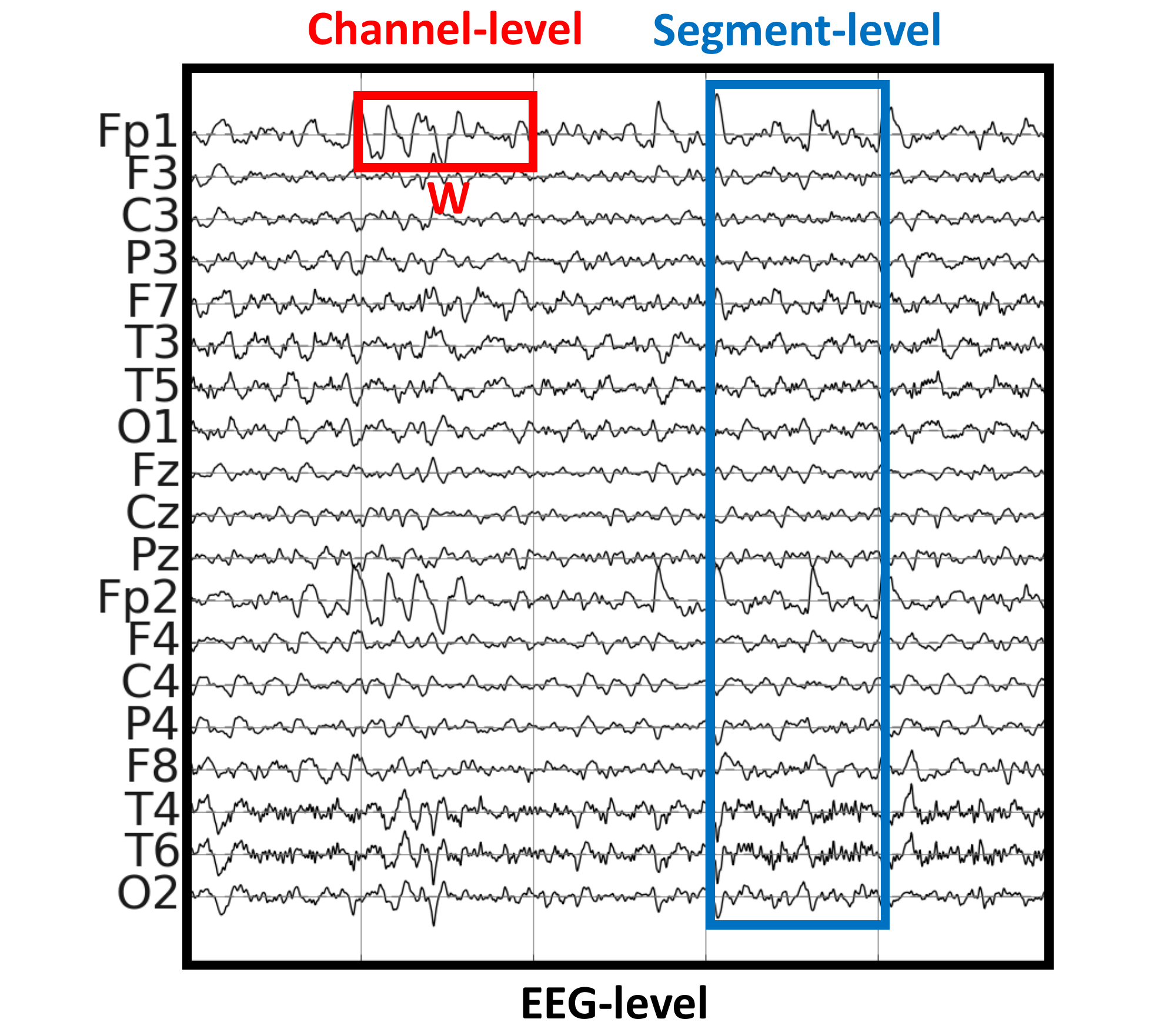}
\caption{The three EEG scales: channel-, segment-, and EEG-level detection.}
\label{fig:EEG_scale}
\end{figurehere}

\subsection{Seizure Detector Pipeline}
\begin{figure*}
\centering
\includegraphics[width=17cm]{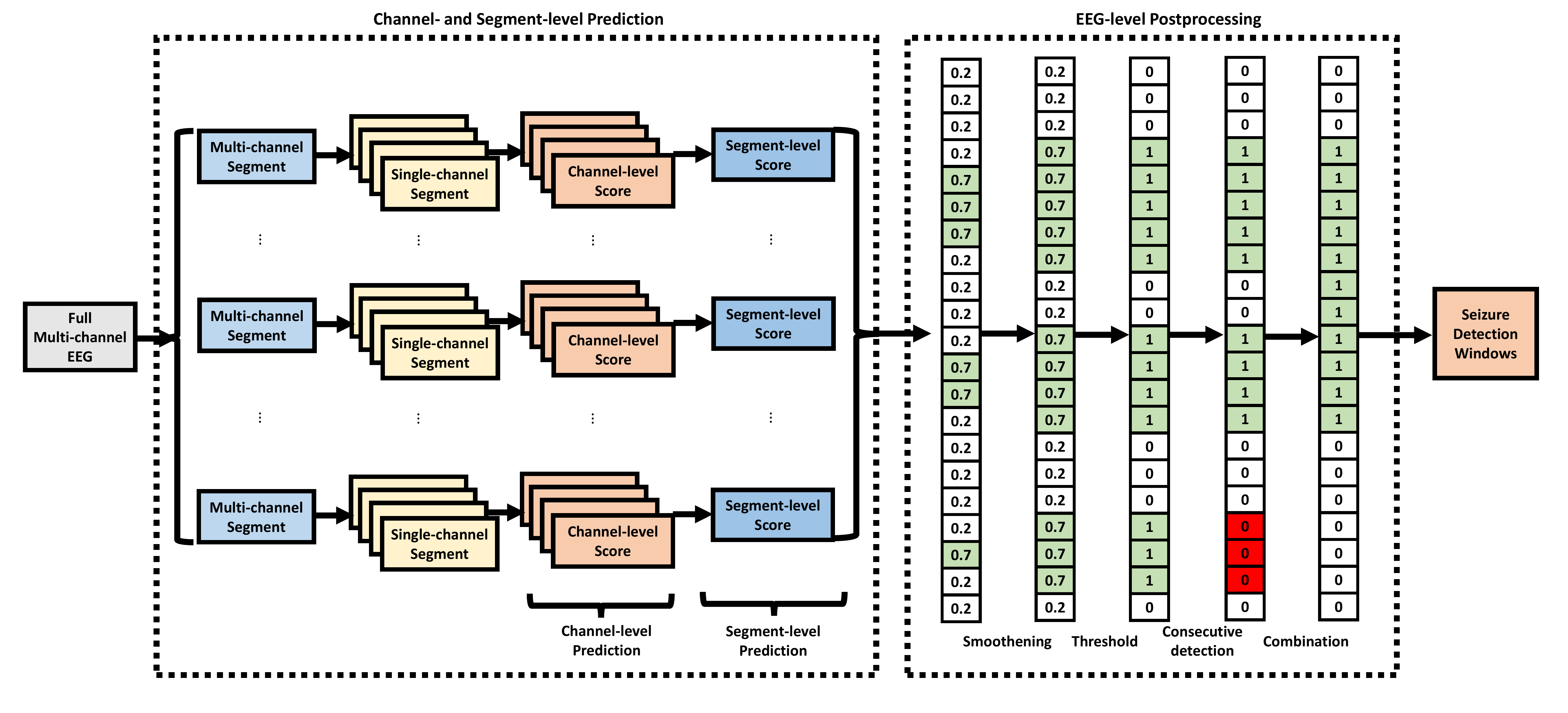}
\caption{The proposed seizure detector pipeline consists of multiple stages of seizure detection at three EEG scales. The EEG is divided into overlapping epochs where we performed channel- and segment-level detection to get a series of segment-level outputs. Next, we applied four postprocessing steps for EEG-level detection. First, we apply smoothening window (e.g., max smoothen window of length 3) to the segment-level output. Second, we implement thresholds (e.g., 0.5) to obtain a series of 0s and 1s. Third, we locate chains of consecutive 1s and replace them with 0s if the chain is less than $N_{c}$ (e.g., $N_{c}=4$) in length. Finally, suppose any two chains of consecutive 1s are within proximity (e.g., 3 epochs). In that case, we combine them into a single detection to prevent many fractured detection windows.}
\label{fig:Classification_model_seizure}
\end{figure*}

We perform seizure detection first at individual channels (channel-level detection), followed by multi-channel segments (segment-level detection). At last, we detect the start and end points of the seizures in the entire multi-channel EEG (EEG-level detection)~\cite{peh2021multi, thomas2021automated, thangavel2021time} (see Figure~\ref{fig:EEG_scale}). The proposed seizure detector is displayed in Figure~\ref{fig:Classification_model_seizure}. The pipeline consists of a channel-level deep learning classifier, a segment-level machine learning classifier, and multiple EEG-level post-processing modules. The seizure detectors are implemented on NVIDIA GeForce GTX1080 GPUs in Keras 2.2.0 and TensorFlow 2.6.0.

\subsection{Channel-level Seizure Detector}
The channel-level seizure detector computes the seizure probability for single-channel EEG segments. The window length $W$ adopted in the literature ranges between 1s to 30s. However, $W=1s$ is too short to capture long-range seizure morphology, while $W=30s$ is too long to capture short seizures. Therefore, we tested window lengths $W \in \{3, 5, 10, 20\}$ seconds. In this study, we deploy three channel-level seizure detectors based on convolutional neural networks (CNN):

\begin{enumerate}[leftmargin=*]
  \item CNN with softmax (SM) loss: CNN-SM
  \item CNN with belief matching (BM) loss: CNN-BM
  \item CNN-transformer with BM loss: CNN-TRF-BM.
\end{enumerate}

\subsubsection{CNN-SM Model}
The CNN-SM model is a CNN with a SM loss function. The input is the raw single-channel signal of length $W\times \text{F}_s$. The architecture contains five convolutional layers with 8, 16, 32, 64, and 128 filters, respectively, with two fully connected layers. To minimize the loss, we applied the Adam optimizer with an initial learning rate equal to $10^{-4}$. The batch size during training is set to 1000. Also, we implemented class weights that are inversely proportional to the class frequency in the training data during training. This allows us to optimize the loss function on an imbalanced dataset without overfitting~\cite{peh2021multi}. Finally, we optimized parameters within the CNN via nested CV on the training data, with an 80:20\% split for training and validation.

\subsubsection{CNN-BM Model}
The CNN-BM model has the same architecture as the CNN-SM model, except that the BM loss replaces the SM loss. The BM loss is shown to yield better uncertainty estimates and generalization performance than the SM loss, an important property required for seizure detection~\cite{joo2020being}. The BM framework is formulated from a Bayesian perspective that views binary classification as distribution matching. The BM loss is defined as:

\begin{equation} \label{eq:BM_Loss}
\mathcal{L}(\textbf{W}) \approx
-\frac{1}{m} \sum_{i=1}^{m} \ell_\text{EB} 
\left(y^{(i)}, \alpha^{\textbf{W}}(\textbf{x}^{(i)}) \right),
\end{equation}

\noindent where $\textbf{x}^{(i)}$ and $y^{(i)}$ is the $i$-th training data and its label, respectively, $m$ is the total number of samples, and $\alpha^{\textbf{W}}=\text{exp} (\textbf{W})$, where $\textbf{W}$ are the weights of the neural network classifier. $\ell_{\text{EB}}(y, \alpha^{\textbf{W}}(\textbf{x}))$ is the evidence lower bound (ELBO)~\cite{joo2020being} and is defined as $\ell_{\text{EB}}(y, \alpha^{\textbf{W}}(\textbf{x})) = \mathbb{E}_{q_{\textbf{z} \, | \, \textbf{x}}^{\textbf{W}}} [ \text{log} \, p (y | \textbf{x}, \textbf{z}) ] - \text{KL} (q_{\textbf{z} \, | \, \textbf{x}}^{\textbf{W}} \, || \, p_{\textbf{z} | \textbf{x}})$, where $\textbf{z}$ is the categorical probability about the label, $ p_{\textbf{z} | \textbf{x}}$ is the target distribution, $q_{\textbf{z}|\textbf{x}}^{\textbf{W}}$ is the approximate distribution, and $\text{KL}$ is the KL-divergence. We refer to~\cite{joo2020being} for more information on the BM loss.

\subsubsection{CNN-TRF-BM Model}
\begin{figure*}
\hfill
\begin{minipage}[b]{0.49\linewidth}
  \centering
  \centerline{\includegraphics[width=7cm]{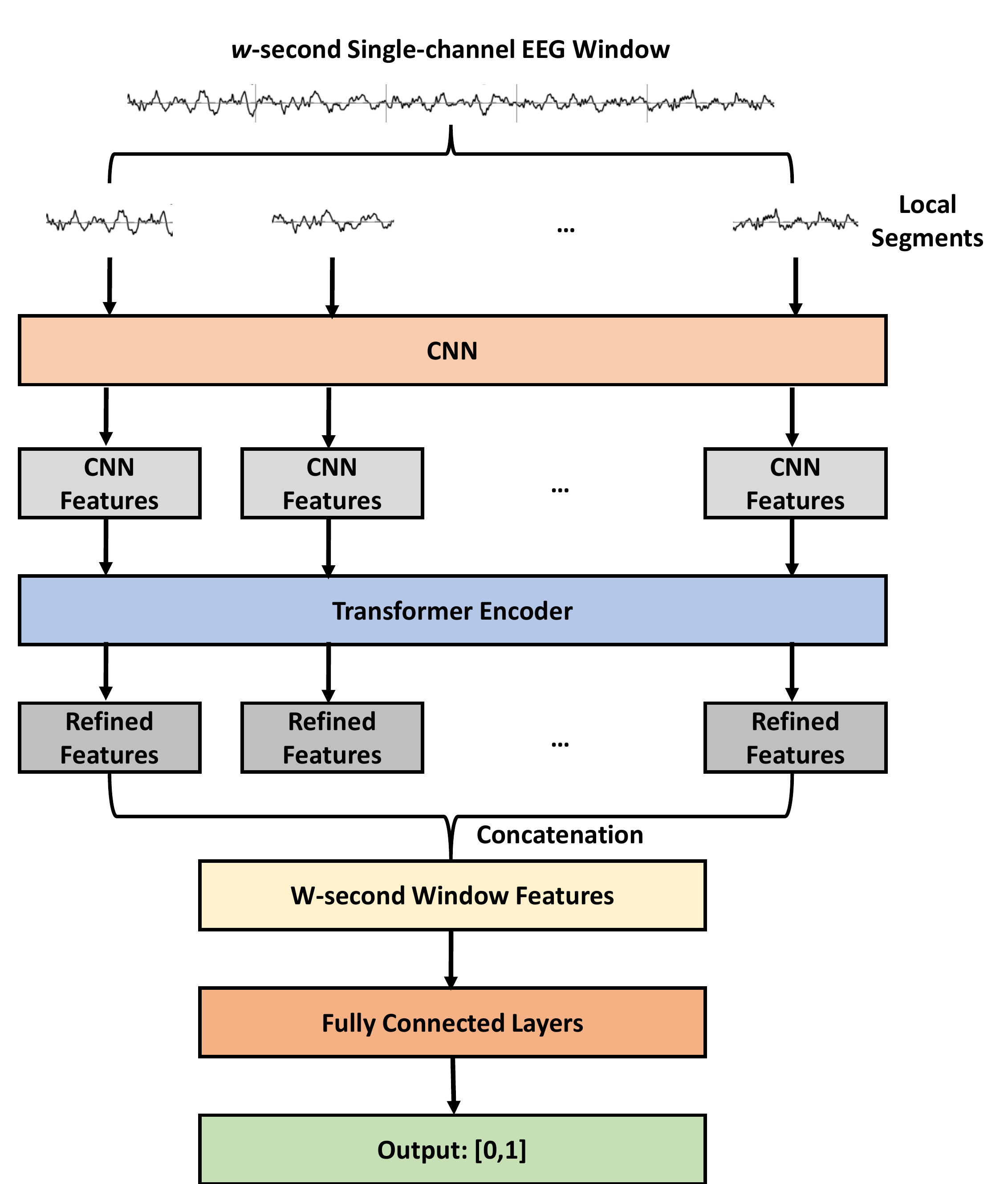}}
  %\vspace{0.5cm}
  \centerline{(a) CNN with transformer encoder.}\medskip
\end{minipage}
\hfill
\begin{minipage}[b]{0.49\linewidth}
  \centering
  \centerline{\includegraphics[width=4.5cm]{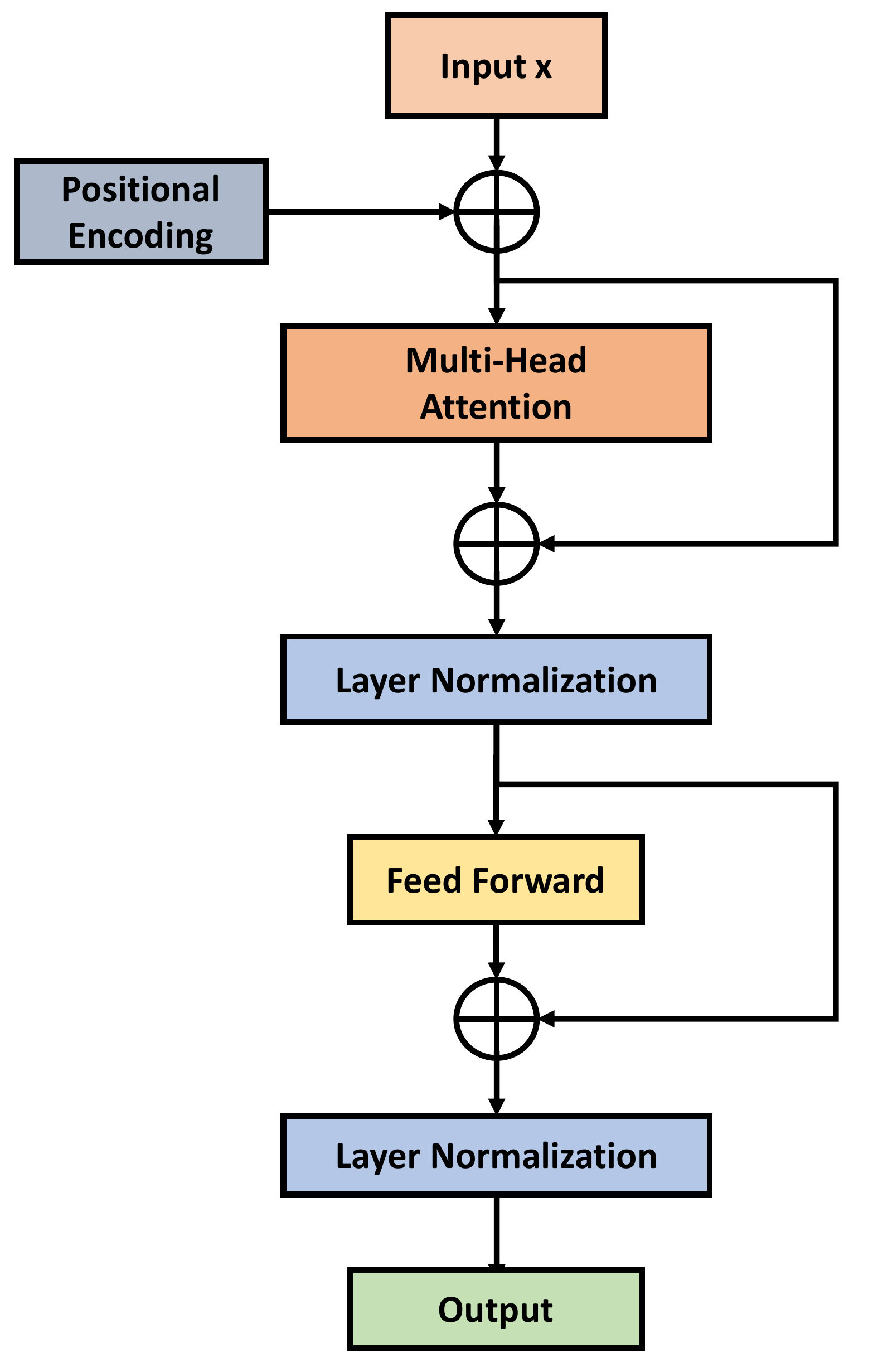}}
  %\vspace{0.5cm}
  \centerline{(b) Transformer encoder.}\medskip
\end{minipage}
\hfill
\vspace{-0.1cm}
\caption{The architecture of the CNN-TRF-BM model in this study. (a) CNN-TRF-BM model, and (b) transformer encoder.}
\label{fig:modified_transformer}
\end{figure*}

The CNN-TRF-BM model contains the CNN and the transformer. The architecture is the same as in the CNN-BM model, but we insert an additional transformer encoder between the final convolutional layer and the flattening layer (see~\cite{peh2022transformer}). We implemented a transformer in tandem with the CNN, as the CNN alone cannot model correlations between distant data points, such as seizure morphologies (see Figure~\ref{fig:modified_transformer}(a) and (b)). The transformer can compensate for this limitation by extracting long-range information from the CNN features. The transformer encoder contains eight heads, and the number of hidden layer neurons in the forward feed network (FFN) is 1024. As input to the transformer, we extract 1s segments with 25\% overlap from the $W$-second single-channel segment.

\subsection{Segment-level Seizure Detector}
Next, we rely on the outputs of the channel-level detectors to detect seizures in multi-channel segments. The channel-level detectors yield seizure probabilities for each EEG channel, which we arrange into regions according to the scalp topology: frontal, central, occipital, and parietal. Besides those four local regions, we also define a ``global" region containing all channels. From each region, we extract seven statistical features: mean, median, standard deviation, maximum value, minimum value, and value at 25\% and 75\% percentile. As there are five regions, we extract $5 \times 7=35$ features. From all channel-level outputs, we compute the normalized histogram features (5 bins, range [0,1]) and include them into the feature set, bringing the total features to $40$.

In the iEEGs, the channel locations are unavailable; hence we cannot group the iEEG channels into local regions. Instead, we replace the four local regions with the global region. In this scenario, only $12$ features are unique, and the remaining ones are duplicates. In any case, the number of segment-level features is $40$, regardless of the number of channels or the availability of the channel locations. This approach ensures that the number of features is consistent during the training and evaluation of any dataset. The features will be the inputs to an XGBoost for training and validation, and we determined the hyperparameters via grid search CV.

\subsection{Channel- and Segment-level Evaluation Metric}
We assess the channel- and segment-level seizure classifiers through the following metrics: accuracy (ACC), balanced accuracy (BAC), sensitivity (SEN), specificity (SPE), F1 score (F1), and expected calibration error (ECE)~\cite{guo2017calibration}. As the seizure and non-seizure classes are imbalanced, we evaluate the results mainly in BAC~\cite{peh2021multi}.

\subsection{EEG-level Seizure Detector}
Finally, we perform seizure detection on full EEGs by determining the start and end time of the seizures, if any. First, we apply a sliding window of length $W$ with an overlap duration $T_o$ to the multi-channel EEG, extracting $n$ multi-channel segments. The overlap duration $T_o$ is set to 1s. Next, we perform segment-level detection on each segment, resulting in $n$ seizure probabilities $P=[p_1, \cdots, p_n]$. Finally, we conduct three post-processing steps to the seizure probability sequence~$P$:

\begin{enumerate}[leftmargin=*]

    \item We apply 1D smoothing filters with an overlap of 1 sample. We tested various filter lengths $K_f$ (3, 5, or 7s) and filter types (mean, median, or max). The smoothing filter removes isolated seizure detections (usually false positives (FPs) such as artifacts) and smoothens regions with significant confidence variations to stabilize the detections.

    \item Next, we perform thresholding to the seizure probabilities to round them to zeros (seizure-free) or ones (seizure). We tested threshold values $\theta \in \{0.1, 0.2, \cdots, 0.8, 0.9\}$.

    \item Then, we identify consecutive ones of length smaller than $N_c$, and replace the 1s with 0s. Selecting a large $N_c$ removes many short detections, leading to fewer FPs and more FNs, as the system may miss short seizures. We tested $N_c \in \{1, 2, \cdots, 19, 20\}$.

\end{enumerate}

Finally, we identify the remaining sequences of consecutive 1s, and determine their start and end time. The final output of the EEG-level seizure detector is the start and end times of the detected seizures.

\subsection{EEG-level Seizure Detection Evaluation Metric}
We assess the accuracy of the detections via EEG-level seizure detection evaluation metric. There are several well-established evaluation metrics, such as epoch-based sampling (EBS)~\cite{shah2021objective}, any-overlap (OVLP)~\cite{shah2021objective}, time-aligned event scoring (TAES)~\cite{shah2021objective}, and increased margin scoring (IMS)~\cite{reus2022automated}. However, these metrics do not accurately reflect the clinical requirement of a seizure detector. Hence, we define a new metric, the minimum overlap evaluation scoring (MOES). In this metric, there needs to be a non-trivial overlap between the detection and the seizure, while it does not need to be perfect.

We elaborate on the limitations of the existing seizure evaluation metrics in the supplementary methods section. In short, OVLP metric considers a detection correct as long as it has a non-zero overlap with the annotation, which is too lenient and leads to overly optimistic results. On the other hand, TAES metric is too strict as it requires a perfect overlap between the detection and annotation, leading to overly pessimistic results.

\subsection{Minimum Overlap Evaluation Scoring (MOES)}
The minimum overlap evaluation scoring (MOES) determines the overlap duration $T_{\text{overlap}}$ between the detection ($T_{\text{detection}} = [d_{\text{start}}, d_{\text{end}}]$) and seizure ($T_{\text{seizure}} = [s_{\text{start}}, s_{\text{end}}]$) window, and vice versa, before deciding if the detection is correct or the seizure is captured. Based on existing literature, only seizures of at least 10s are annotated typically~\cite{afra2008duration}. Therefore, the minimum overlap duration of the detection(s) with the seizure should be 10s. However, these criteria do not account for the duration of the seizure or the detection. Therefore, even if the detection correctly detected over 10s of a seizure, the system should be penalised if the majority of the detection did not capture any seizure. To resolve this, we compute the detection overlap (DOL) and the seizure overlap (SOL), which measures the fraction of the detection that overlaps with any seizures, and vice versa, as:

\begin{equation} \label{eq:DOL}
\text{DOL}_{i} = \frac{\sum_{s}{T_{\text{overlap},s,i}}}{d_{\text{end},i} - d_{\text{start},i}},
\end{equation}

\begin{equation} \label{eq:SOL}
\text{SOL}_{j} = \frac{\sum_{d}{T_{\text{overlap},d,j}}}{s_{\text{end},j} - s_{\text{start},j}},
\end{equation}

\noindent where $i$ and $j$ is the index of a detection and a seizure, respectively, $\sum{T_{\text{overlap},s,i}}$ is the sum of all the overlaps with any seizures with detection $i$, and $\sum{T_{\text{overlap},d,j}}$ is the sum of all the overlaps with any seizures with seizure $j$.

In this study, we set a minimum DOL and SOL of 0.3 (30\%) to ensure that a significant portion of the detection overlaps with the seizures and vice versa. In OVLP metric, the DOL is set to be 0+\%, while in TAES it is 100\%. The first option is too lenient in practice, while the latter is too strict.

A high DOL implies that the detection overlaps well with the seizure(s). Meanwhile, a high SOL indicates that the seizure is well captured by the detection(s). If the DOL is low, the detection should be discarded and treated as a false positive (FP). Similarly, if the SOL is low, the seizure should be treated as a false negative (FN). More details on how MOES approaches different detection cases are elaborated in the extended version of the paper.

\begin{figure*}[htp]
\centering
\includegraphics[width=16cm]{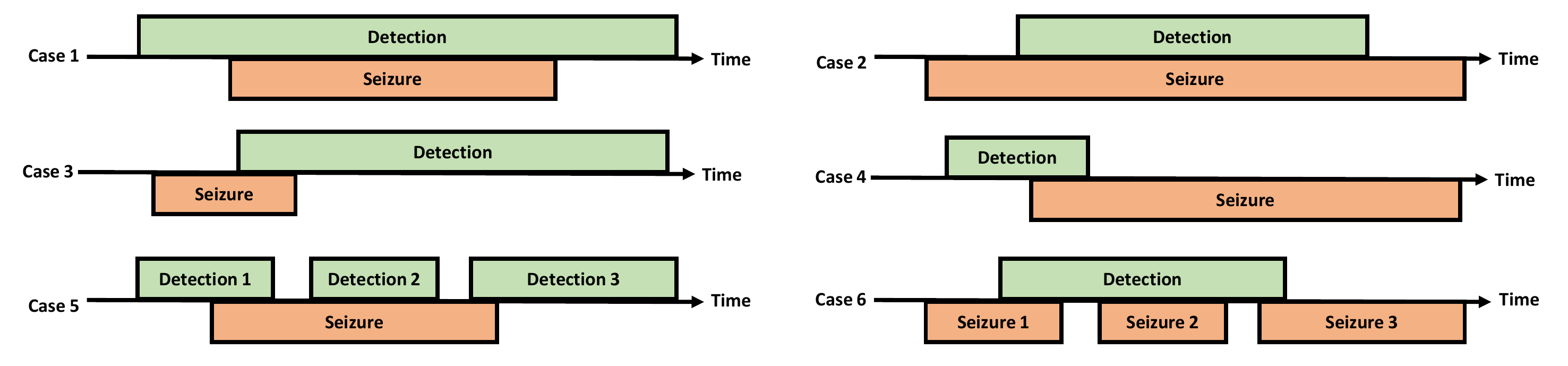}
\vspace{-0.3cm}
\caption{Different cases of seizure detection that may be encountered.}
\label{fig:MOES_cases}
\end{figure*}

This approach allows us to consider different cases (see Figure~\ref{fig:MOES_cases}):

\begin{itemize}[leftmargin=*]

    \item Case 1: The detection window encapsulates the seizure window almost perfectly. Therefore, the $\text{SOL}_{j}=1$, while $\text{DOL}_{i}>0.3$ (close to 1). In this case, the $\text{seizure}_{j}$ is a TP as the $\text{detection}_{i}$ is correct.

    \item Case 2: The detection window is encapsulated by the seizure window almost perfectly. Therefore, the $\text{DOL}_{i}=1$, while $\text{SOL}_{j}>0.3$ (close to 1). In this case, the $\text{seizure}_{j}$ is a TP as the $\text{detection}_{i}$ is correct.

    \item Case 3: The detection window overlaps with the seizure window, however, the detection window protrudes the seizure window by a significant margin. In this case, while the $\text{SOL}_{j}$ can be greater than 0.3, the $\text{DOL}_{i}$ is low (less than 0.3). As $\text{DOL}_{i}<0.3$, we consider the $\text{detection}_{i}$ as a false alarm (FP). As a result, the $\text{seizure}_{j}$ is considered as a FN.

    \item Case 4: The seizure window overlaps the detection window, however, the seizure window protrudes the detection window by a significant margin. In this case, while the $\text{DOL}_{i}$ can be greater than 0.3, the $\text{SOL}_{j}$ is low (less than 0.3). As $\text{SOL}_{j}<0.3$, we consider the $\text{seizure}_{j}$ as missed (FN). As a result, the $\text{detection}_{i}$ is considered as a FP.

    \item Case 5: Multiple detection windows (1, 2, 3) overlap with the annotated seizure. The majority of the $\text{seizure}_{j}$ is detected, hence the $\text{SOL}_{j}$ is high (greater than 0.3). However, the $\text{DOL}_{i}$ vary for each detection, though all of them clipped the seizure to a certain extent. 

        \begin{enumerate}
            \item $\text{Detection}_{1}$ would have $\text{DOL}_{1} \approx 0.5$, hence it is a correct detection.
            \item $\text{Detection}_{2}$ would have $\text{DOL}_{2} = 1$, hence it is a correct detection.
            \item $\text{Detection}_{3}$ would have $\text{DOL}_{3} < 0.3$, hence it is a false detection.
        \end{enumerate}

    As $\text{SOL}_{j}>0.3$, we consider the $\text{seizure}_{j}$ as captured, hence a true positive. Meanwhile, the $\text{detection}_{1}$ and $\text{detection}_{2}$ are correct (TP) and $\text{detection}_{3}$ is considered as a FP.

    \item Case 6: Multiple seizure windows (1, 2, 3) overlap with a detection window. The majority of the $\text{detection}_{i}$ had capture seizures, hence the $\text{DOL}_{i}$ is high (greater than 0.3). However, the $\text{SOL}_{j}$ vary for each seizure, though all of them clipped the detection to a certain extent. 

        \begin{enumerate}
            \item $\text{Seizure}_{1}$ would have $\text{SOL}_{1} \approx 0.5$, hence the seizure is detected well.
            \item $\text{Seizure}_{2}$ would have $\text{SOL}_{2} = 1$, hence the seizure is detected well.
            \item $\text{Seizure}_{3}$ would have $\text{SOL}_{3} < 0.3$, hence the seizure is not detected.
        \end{enumerate}

    As $\text{DOL}_{i}>0.3$, we consider the $\text{detection}_{j}$ as correct, hence it is not a FP. Meanwhile, the $\text{seizure}_{1}$ and $\text{seizure}_{2}$ are one TP each and $\text{detection}_{3}$ is missed and is considered as a false negative.

\end{itemize}

In other words, the detections and seizures are analyzed separately. By investigating the seizures, we can compute the TP. Firstly, we check what detections overlap with the $\text{seizure}_{j}$. For the seizure to be considered a TP, the following two conditions must be met simultaneously:

\begin{enumerate}[leftmargin=*]
    \item $\text{SOL}_{j} \geq 0.3$.
    \item $\text{DOL}_{i} \geq 0.3$ for all detections overlapping with $\text{seizure}_{j}$. 
\end{enumerate}

The first condition implies that $\text{seizure}_{j}$ is sufficiently captured by one or more detections. The second condition makes sure that each of those detections sufficiently covers a seizure ($\text{seizure}_{j}$ and potentially also other seizures). If both conditions are met simultaneously, the seizure is accurately detected and is a TP. Otherwise, the seizure is missed, and it is a FN. Indeed, imagine that 2 detections on the left and right of the seizure last very long and cover together almost the entire EEG. Then the entire EEG is covered by detections, and the seizure would not be properly detected. The two detections could have very low DOL. Consequently, the seizure in that EEG will be considered a FN instead of a TP. 

Next, by investigating the detections, we compute the FPs. We first determine what seizures overlap with $\text{detection}_{i}$. The detection is considered a FP, as long as any of the following two conditions are met:

\begin{enumerate}[leftmargin=*]
    \item $\text{DOL}_{i} < 0.3$
    \item $\text{SOL}_{j} < 0.3$ for all seizures overlapping with $\text{detection}_{i}$. 
\end{enumerate}

Note that it is important to compute TPs from the perspective of the seizure. Indeed, multiple detections may overlap with the same seizure (see Figure~\ref{fig:MOES_cases}, case 6). However, as there is only one seizure event, we only can have one TP or one FN associated with a seizure event. Therefore, we need to compute the TPs from the perspective of the seizures. Computing TP from the perspective of the detection windows may result in multiple TPs for a single seizure event, which is undesirable.

Finally, the detection may start earlier or later than the annotated seizure. We compute the detection offset as:

\begin{equation} \label{eq:Detection_Offset_Our}
T_{\text{offset}}= d_{\text{start}} - s_{\text{start}} + W,
\end{equation}

\noindent where $W$ are the duration of the window length, $d_{\text{start}}$ is the start time of the detection, $s_{\text{start}}$ is the start time of the annotated seizure. We added $W$ in the offset as we require a minimum window of length $W$ to detect seizures. To more accurately detect the onset of a seizure, one may slide the window in smaller steps around the onset of a detection. However, this goes beyond the scope of this work, as we are mainly interested in detecting seizures, irrespective of their onset times.

\subsection{EEG-level Seizure Detection Performance Metrics}
We measure the performance of EEG-level seizure detection with sensitivity (SEN), precision (PRE), and false positive per hour (FPR/h). We report both the average FPR/h (aFPR/h) and median FPR/h (mFPR/h). We mainly focus on the mFPR/h in this study as they are more robust to outliers compared to the average.

%%%%%%%%%%%%%%%%%%%%%%%%%%%%%%%%%%%%%%%%%%
% Results
\section{Results}

\begin{figure*}[htp]
\centering
\hspace{-1cm}
\includegraphics[width=18cm]{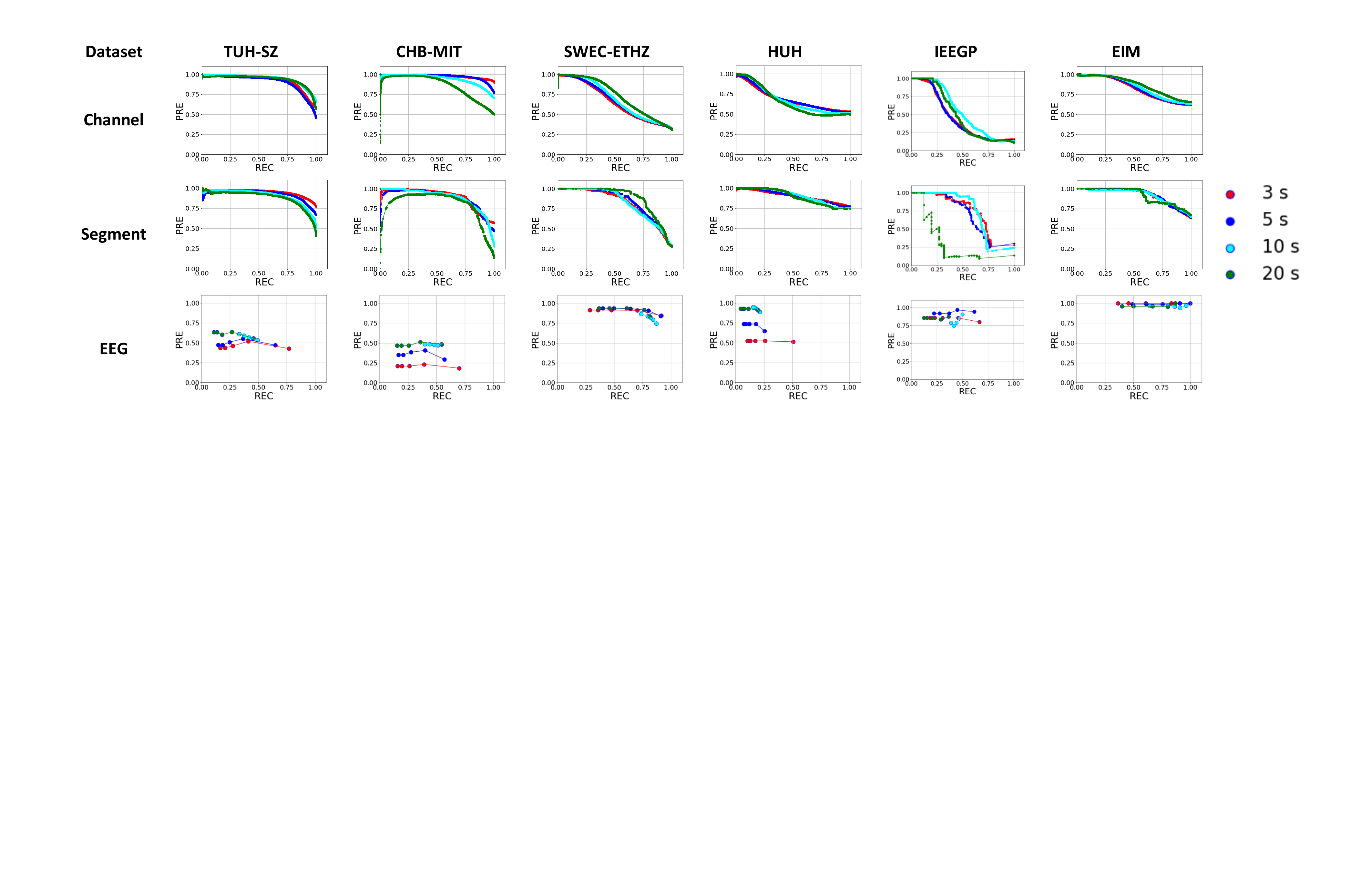}
\vspace*{-7cm}
\caption{The precision-recall (PR) curves of the channel-, segment-, and EEG-level seizure detection computed with the CNN-TRF-BM-based seizure detector across different datasets. The EEG-level PR curves are generated by varying the threshold $\theta$ in the EEG-level postprocessing step and computing the precision (PRE) and recall (REC) at each threshold with MOES.}
\label{fig:precision_recall_curve}
\end{figure*}

\subsection{Channel-level Seizure Detection}
We performed channel-level seizure detection with three channel-level detectors: CNN-SM, CNN-BM, and CNN-TRF-BM. We summarized the results in Table~\ref{tab:LOSO_Channel}. In addition, all the precision-recall (PR) curves can be found in Figure~\ref{fig:precision_recall_curve}.

On the TUH-SZ dataset, the proposed channel-level detectors achieve high BAC, SEN, and SPE across all window lengths. Moreover, the ECE improved for all window lengths (except for 3s) when the SM loss is replaced with the BM loss (CNN-SM against CNN-BM). However, the ECE is slightly larger for the CNN-TRF-BM model. The performance peaks at a $W=20s$ for all three models. Overall, the CNN-TRF-BM model attained the best results, followed by the CNN-BM and the CNN-SM model. As the channel-level detector attains good results on the TUH-SZ dataset, using it as the primary training dataset seems promising.

Next, we assessed the channel-level detector, trained on the TUH-SZ dataset, on the five EEG datasets. The detectors achieve high BACs on the CHB-MIT, SWEC-ETHZ, and EIM datasets, but yield poor BACs on the HUH and IEEGP datasets. For those datasets, seizures have only been annotated on the level of segments instead of channels; therefore, it is impossible to assess the channel detector reliably. Without channel-level annotations, we must assume that all channels within a multi-channel segment contain seizures. However, this is unlikely as seizures sometimes only occur in certain regions. In particular, focal seizures occur only in one hemisphere or at a few electrodes. Consequently, channels that do not exhibit seizures may be mislabelled as ``seizures", leading to errors during training and testing. However, segment-level and EEG-level detection results are reliable for those datasets.

\subsection{Segment-level Seizure Detection}
Next, we performed segment-level seizure detection using the outputs from the three channel-level detectors. The segment-level detection results on the six EEG datasets are displayed in Table~\ref{tab:LOSO_Segment}.

On the TUH-SZ dataset, the proposed segment-level detectors achieve high BAC, SEN, and SPE across all window lengths, similarly to the channel-level results. However, the ECE reported at the segment-level is much greater than the channel-level counterparts, as the segment-level detector model does not minimize ECE. Similarly, the performance peaks at a $W=20s$. Again, the CNN-TRF-BM model outshines the other two models.

Next, we evaluated the segment-level seizure detector on the other five datasets. We obtained excellent performance on all the datasets at various window lengths, except for the HUH dataset. The segment-level detectors obtain high BACs on the IEEGP dataset, even when the channel-level results on this dataset are not satisfactory.

Overall, the performance peaks at different window lengths across the six datasets. This might be due to the discrepancy in seizure types, patient types, and patient age groups across the different datasets. For instance, for datasets with many short seizures, one should deploy a window length of 3s as it can capture shorter seizures, while a window length of 20s would be suboptimal.

\begin{table*}
\centering
\tbl{Channel-level seizure detection results for different CNN models across six EEG datasets. \label{tab:LOSO_Channel}}
{
\scalebox{0.68}{
\begin{tabular}{|c|c|cccccc|cccccc|cccccc|}

\hline \hline
\multirow{2}{*}{\textbf{Dataset}} & \multirow{2}{*}{\textbf{W}} & \multicolumn{6}{c|}{\textbf{CNN-SM}} & \multicolumn{6}{c|}{\textbf{CNN-BM}} & \multicolumn{6}{c|}{\textbf{CNN-TRF-BM}} \\ 
\cline{3-20}
 & & \textbf{ECE} & \textbf{ACC} & \textbf{BAC} & \textbf{SEN} & \textbf{SPE} & \textbf{F1} & \textbf{ECE} & \textbf{ACC} & \textbf{BAC} & \textbf{SEN} & \textbf{SPE} & \textbf{F1} & \textbf{ECE} & \textbf{ACC} & \textbf{BAC} & \textbf{SEN} & \textbf{SPE} & \textbf{F1} \\ 
\hline \hline

\multirow{4}{*}{\begin{tabular}[c]{@{}c@{}}TUH-SZ\\Scalp EEG\\ Adult\end{tabular}} & 3 & 0.043 & 0.824 & 0.832 & 0.808 & 0.855 & 0.827 & 0.046 & 0.837 & 0.842 & 0.827 & 0.856 & 0.839 & 0.052 & 0.824 & 0.832 & 0.773 & 0.89 & 0.826 \\
 & 5 & 0.043 & 0.84 & 0.836 & 0.769 & 0.902 & 0.84 & 0.035 & 0.845 & 0.842 & 0.862 & 0.821 & 0.848 & 0.03 & 0.85 & 0.83 & 0.767 & 0.892 & 0.849 \\
 & 10 & 0.044 & 0.815 & 0.826 & 0.809 & 0.844 & 0.821 & 0.021 & 0.848 & 0.844 & 0.78 & 0.908 & 0.848 & 0.056 & 0.772 & 0.76 & 0.868 & 0.653 & 0.758 \\
 & 20 & \cellcolor[gray]{0.9}{0.044} & \cellcolor[gray]{0.9}{0.836} & \cellcolor[gray]{0.9}{0.845} & \cellcolor[gray]{0.9}{0.812} & \cellcolor[gray]{0.9}{0.877} & \cellcolor[gray]{0.9}{0.837} & \cellcolor[gray]{0.9}{0.027} & \cellcolor[gray]{0.9}{0.845} & \cellcolor[gray]{0.9}{0.851} & \cellcolor[gray]{0.9}{0.834} & \cellcolor[gray]{0.9}{0.868} & \cellcolor[gray]{0.9}{0.846} & \cellcolor[gray]{0.9}{0.033} & \cellcolor[gray]{0.9}{0.852} & \cellcolor[gray]{0.9}{0.858} & \cellcolor[gray]{0.9}{0.828} & \cellcolor[gray]{0.9}{0.889} & \cellcolor[gray]{0.9}{0.853} \\ 
\hline

\multirow{4}{*}{\begin{tabular}[c]{@{}c@{}}HUH\\Scalp EEG\\ Neonatal\end{tabular}} & 3 & 0.259 & 0.506 & 0.491 & 0.187 & 0.794 & 0.454 & 0.399 & 0.403 & 0.403 & 0.249 & 0.903 & 0.496 & 0.408 & 0.4 & 0.4 & 0.245 & 0.902 & 0.492 \\
 & 5 & 0.28 & 0.532 & 0.511 & 0.12 & 0.902 & 0.445 & 0.481 & 0.354 & 0.354 & 0.168 & 0.957 & 0.423 & \cellcolor[gray]{0.9}{0.377} & \cellcolor[gray]{0.9}{0.427} & \cellcolor[gray]{0.9}{0.427} & \cellcolor[gray]{0.9}{0.289} & \cellcolor[gray]{0.9}{0.879} & \cellcolor[gray]{0.9}{0.526} \\
 & 10 & 0.228 & 0.527 & 0.507 & 0.217 & 0.796 & 0.482 & \cellcolor[gray]{0.9}{0.403} & \cellcolor[gray]{0.9}{0.417} & \cellcolor[gray]{0.9}{0.417} & \cellcolor[gray]{0.9}{0.264} & \cellcolor[gray]{0.9}{0.912} & \cellcolor[gray]{0.9}{0.511} & 0.508 & 0.358 & 0.358 & 0.168 & 0.974 & 0.423 \\
 & 20 & \cellcolor[gray]{0.9}{0.271} & \cellcolor[gray]{0.9}{0.574} & \cellcolor[gray]{0.9}{0.534} & \cellcolor[gray]{0.9}{0.131} & \cellcolor[gray]{0.9}{0.937} & \cellcolor[gray]{0.9}{0.485} & 0.457 & 0.385 & 0.385 & 0.211 & 0.952 & 0.464 & 0.527 & 0.343 & 0.343 & 0.145 & 0.986 & 0.403 \\ 
\hline

\multirow{4}{*}{\begin{tabular}[c]{@{}c@{}}CHB-MIT\\Scalp EEG\\ Paediatric\end{tabular}} & 3 & 0.259 & 0.617 & 0.756 & 0.569 & 0.942 & 0.649 & 0.269 & 0.568 & 0.74 & 0.51 & 0.97 & 0.601 & 0.25 & 0.582 & 0.747 & 0.528 & 0.966 & 0.617 \\
 & 5 & 0.181 & 0.669 & 0.763 & 0.56 & 0.966 & 0.668 & 0.205 & 0.62 & 0.739 & 0.494 & 0.984 & 0.616 & \cellcolor[gray]{0.9}{0.095} & \cellcolor[gray]{0.9}{0.742} & \cellcolor[gray]{0.9}{0.808} & \cellcolor[gray]{0.9}{0.666} & \cellcolor[gray]{0.9}{0.95} & \cellcolor[gray]{0.9}{0.755} \\
 & 10 & \cellcolor[gray]{0.9}{0.126} & \cellcolor[gray]{0.9}{0.786} & \cellcolor[gray]{0.9}{0.816} & \cellcolor[gray]{0.9}{0.743} & \cellcolor[gray]{0.9}{0.889} & \cellcolor[gray]{0.9}{0.79} & 0.137 & 0.724 & 0.782 & 0.635 & 0.928 & 0.733 & 0.205 & 0.663 & 0.748 & 0.515 & 0.981 & 0.649 \\
 & 20 & 0.129 & 0.777 & 0.78 & 0.592 & 0.969 & 0.758 & \cellcolor[gray]{0.9}{0.141} & \cellcolor[gray]{0.9}{0.777} & \cellcolor[gray]{0.9}{0.782} & \cellcolor[gray]{0.9}{0.606} & \cellcolor[gray]{0.9}{0.959} & \cellcolor[gray]{0.9}{0.765} & 0.153 & 0.755 & 0.756 & 0.534 & 0.978 & 0.733 \\ 
\hline

\multirow{4}{*}{\begin{tabular}[c]{@{}c@{}}SWEC-ETHZ\\ iEEG\\ Adult\end{tabular}} & 3 & 0.069 & 0.803 & 0.721 & 0.56 & 0.882 & 0.804 & 0.127 & 0.814 & 0.725 & 0.557 & 0.892 & 0.813 & 0.107 & 0.814 & 0.726 & 0.56 & 0.891 & 0.814 \\
 & 5 & 0.066 & 0.828 & 0.718 & 0.502 & 0.935 & 0.819 & 0.108 & 0.834 & 0.723 & 0.514 & 0.933 & 0.826 & 0.097 & 0.798 & 0.73 & 0.614 & 0.847 & 0.805 \\
 & 10 & 0.084 & 0.772 & 0.726 & 0.648 & 0.805 & 0.785 & 0.112 & 0.805 & 0.74 & 0.628 & 0.853 & 0.812 & 0.094 & 0.844 & 0.737 & 0.535 & 0.939 & 0.837 \\
 & 20 & \cellcolor[gray]{0.9}{0.074} & \cellcolor[gray]{0.9}{0.837} & \cellcolor[gray]{0.9}{0.781} & \cellcolor[gray]{0.9}{0.615} & \cellcolor[gray]{0.9}{0.914} & \cellcolor[gray]{0.9}{0.836} & \cellcolor[gray]{0.9}{0.099} & \cellcolor[gray]{0.9}{0.827} & \cellcolor[gray]{0.9}{0.777} & \cellcolor[gray]{0.9}{0.635} & \cellcolor[gray]{0.9}{0.89} & \cellcolor[gray]{0.9}{0.83} & \cellcolor[gray]{0.9}{0.12} & \cellcolor[gray]{0.9}{0.863} & \cellcolor[gray]{0.9}{0.79} & \cellcolor[gray]{0.9}{0.594} & \cellcolor[gray]{0.9}{0.953} & \cellcolor[gray]{0.9}{0.857} \\ 
\hline

\multirow{4}{*}{\begin{tabular}[c]{@{}c@{}}IEEGP\\ iEEG\\ Adult\end{tabular}} & 3 & 0.358 & 0.536 & 0.536 & 0.453 & 0.952 & 0.613 & 0.346 & 0.533 & 0.533 & 0.444 & 0.975 & 0.608 & \cellcolor[gray]{0.9}{0.351} & \cellcolor[gray]{0.9}{0.532} & \cellcolor[gray]{0.9}{0.532} & \cellcolor[gray]{0.9}{0.445} & \cellcolor[gray]{0.9}{0.968} & \cellcolor[gray]{0.9}{0.606} \\
 & 5 & 0.417 & 0.512 & 0.512 & 0.416 & 0.991 & 0.578 & 0.398 & 0.502 & 0.502 & 0.404 & 0.993 & 0.567 & 0.317 & 0.553 & 0.553 & 0.473 & 0.95 & 0.626 \\
 & 10 & \cellcolor[gray]{0.9}{0.317} & \cellcolor[gray]{0.9}{0.574} & \cellcolor[gray]{0.9}{0.574} & \cellcolor[gray]{0.9}{0.508} & \cellcolor[gray]{0.9}{0.9} & \cellcolor[gray]{0.9}{0.651} & 0.352 & 0.562 & 0.562 & 0.479 & 0.976 & 0.631 & 0.386 & 0.523 & 0.523 & 0.428 & 0.998 & 0.59 \\
 & 20 & 0.465 & 0.531 & 0.531 & 0.438 & 0.995 & 0.592 & \cellcolor[gray]{0.9}{0.406} & \cellcolor[gray]{0.9}{0.546} & \cellcolor[gray]{0.9}{0.546} & \cellcolor[gray]{0.9}{0.458} & \cellcolor[gray]{0.9}{0.985} & \cellcolor[gray]{0.9}{0.614} & 0.433 & 0.505 & 0.505 & 0.407 & 0.999 & 0.561 \\ 
\hline

\multirow{4}{*}{\begin{tabular}[c]{@{}c@{}}EIM\\ iEEG\\ Adult\end{tabular}} & 3 & 0.201 & 0.653 & 0.662 & 0.583 & 0.741 & 0.643 & 0.128 & 0.658 & 0.669 & 0.579 & 0.759 & 0.649 & 0.144 & 0.659 & 0.666 & 0.588 & 0.745 & 0.651 \\
 & 5 & 0.205 & 0.65 & 0.684 & 0.52 & 0.848 & 0.633 & 0.135 & 0.652 & 0.687 & 0.518 & 0.855 & 0.638 & 0.154 & 0.66 & 0.653 & 0.626 & 0.679 & 0.653 \\
 & 10 & 0.207 & 0.659 & 0.641 & 0.658 & 0.624 & 0.65 & 0.154 & 0.666 & 0.663 & 0.622 & 0.704 & 0.66 & 0.155 & 0.665 & 0.701 & 0.536 & 0.866 & 0.653 \\
 & 20 & \cellcolor[gray]{0.9}{0.221} & \cellcolor[gray]{0.9}{0.671} & \cellcolor[gray]{0.9}{0.703} & \cellcolor[gray]{0.9}{0.57} & \cellcolor[gray]{0.9}{0.835} & \cellcolor[gray]{0.9}{0.662} & \cellcolor[gray]{0.9}{0.15} & \cellcolor[gray]{0.9}{0.674} & \cellcolor[gray]{0.9}{0.695} & \cellcolor[gray]{0.9}{0.594} & \cellcolor[gray]{0.9}{0.796} & \cellcolor[gray]{0.9}{0.669} & \cellcolor[gray]{0.9}{0.139} & \cellcolor[gray]{0.9}{0.667} & \cellcolor[gray]{0.9}{0.716} & \cellcolor[gray]{0.9}{0.541} & \cellcolor[gray]{0.9}{0.89} & \cellcolor[gray]{0.9}{0.658} \\ 
\hline \hline

\end{tabular}
}}
\end{table*}

\begin{table*}
\centering
\tbl{Segment-level seizure detection results for different CNN models across six EEG datasets. \label{tab:LOSO_Segment}}
{
\scalebox{0.68}{
\centering
\begin{tabular}{|c|c|cccccc|cccccc|cccccc|} 

\hline \hline
\multirow{2}{*}{\textbf{Dataset}} & \multirow{2}{*}{\textbf{W}} & \multicolumn{6}{c|}{\textbf{CNN-SM}} & \multicolumn{6}{c|}{\textbf{CNN-BM}} & \multicolumn{6}{c|}{\textbf{CNN-TRF-BM}} \\ 
\cline{3-20}
 & & \textbf{ECE} & \textbf{ACC} & \textbf{BAC} & \textbf{SEN} & \textbf{SPE} & \textbf{F1} & \textbf{ECE} & \textbf{ACC} & \textbf{BAC} & \textbf{SEN} & \textbf{SPE} & \textbf{F1} & \textbf{ECE} & \textbf{ACC} & \textbf{BAC} & \textbf{SEN} & \textbf{SPE} & \textbf{F1} \\ 
\hline \hline

\multirow{4}{*}{\begin{tabular}[c]{@{}c@{}}TUH-SZ\\Scalp EEG\\ Adult\end{tabular}} & 3 & 0.051 & 0.818 & 0.736 & 0.888 & 0.584 & 0.817 & 0.027 & 0.820 & 0.733 & 0.901 & 0.565 & 0.816 & 0.262 & 0.823 & 0.751 & 0.885 & 0.616 & 0.824 \\
 & 5 & 0.036 & 0.804 & 0.779 & 0.856 & 0.702 & 0.804 & 0.033 & 0.810 & 0.789 & 0.856 & 0.722 & 0.811 & 0.248 & 0.814 & 0.794 & 0.856 & 0.732 & 0.815 \\
 & 10 & 0.039 & 0.815 & 0.817 & 0.783 & 0.850 & 0.815 & \cellcolor[gray]{0.9}{0.031} & \cellcolor[gray]{0.9}{0.833} & \cellcolor[gray]{0.9}{0.833} & \cellcolor[gray]{0.9}{0.815} & \cellcolor[gray]{0.9}{0.852} & \cellcolor[gray]{0.9}{0.833} & 0.027 & 0.832 & 0.831 & 0.800 & 0.862 & 0.831 \\
 & 20 & \cellcolor[gray]{0.9}{0.268} & \cellcolor[gray]{0.9}{0.833} & \cellcolor[gray]{0.9}{0.823} & \cellcolor[gray]{0.9}{0.766} & \cellcolor[gray]{0.9}{0.881} & \cellcolor[gray]{0.9}{0.833} & 0.031 & 0.841 & 0.829 & 0.771 & 0.888 & 0.841 & \cellcolor[gray]{0.9}{0.251} & \cellcolor[gray]{0.9}{0.856} & \cellcolor[gray]{0.9}{0.846} & \cellcolor[gray]{0.9}{0.795} & \cellcolor[gray]{0.9}{0.897} & \cellcolor[gray]{0.9}{0.855} \\ 
\hline

\multirow{4}{*}{\begin{tabular}[c]{@{}c@{}}HUH\\Scalp EEG\\ Neonatal\end{tabular}} & 3 & 0.193 & 0.514 & 0.510 & 0.514 & 0.507 & 0.534 & \cellcolor[gray]{0.9}{0.130} & \cellcolor[gray]{0.9}{0.776} & \cellcolor[gray]{0.9}{0.776} & \cellcolor[gray]{0.9}{0.746} & \cellcolor[gray]{0.9}{0.926} & \cellcolor[gray]{0.9}{0.803} & \cellcolor[gray]{0.9}{0.259} & \cellcolor[gray]{0.9}{0.614} & \cellcolor[gray]{0.9}{0.614} & \cellcolor[gray]{0.9}{0.577} & \cellcolor[gray]{0.9}{0.735} & \cellcolor[gray]{0.9}{0.710} \\
 & 5 & 0.200 & 0.470 & 0.545 & 0.376 & 0.714 & 0.471 & 0.232 & 0.746 & 0.746 & 0.709 & 0.932 & 0.784 & 0.303 & 0.533 & 0.533 & 0.429 & 0.869 & 0.618 \\
 & 10 & 0.353 & 0.407 & 0.575 & 0.192 & 0.957 & 0.349 & 0.366 & 0.651 & 0.651 & 0.581 & 1 & 0.695 & 0.467 & 0.455 & 0.455 & 0.292 & 0.984 & 0.514 \\
 & 20 & \cellcolor[gray]{0.9}{0.357} & \cellcolor[gray]{0.9}{0.413} & \cellcolor[gray]{0.9}{0.575} & \cellcolor[gray]{0.9}{0.183} & \cellcolor[gray]{0.9}{0.968} & \cellcolor[gray]{0.9}{0.349} & 0.414 & 0.628 & 0.628 & 0.533 & 0.817 & 0.691 & 0.444 & 0.426 & 0.426 & 0.251 & 0.994 & 0.483 \\ 
\hline

\multirow{4}{*}{\begin{tabular}[c]{@{}c@{}}CHB-MIT\\Scalp EEG\\ Paediatric\end{tabular}} & 3 & 0.122 & 0.789 & 0.801 & 0.804 & 0.798 & 0.789 & 0.117 & 0.798 & 0.811 & 0.819 & 0.804 & 0.801 & 0.258 & 0.833 & 0.847 & 0.808 & 0.886 & 0.837 \\
 & 5 & 0.105 & 0.814 & 0.824 & 0.762 & 0.887 & 0.808 & 0.126 & 0.811 & 0.816 & 0.700 & 0.932 & 0.808 & 0.256 & 0.822 & 0.824 & 0.715 & 0.932 & 0.819 \\
 & 10 & \cellcolor[gray]{0.9}{0.118} & \cellcolor[gray]{0.9}{0.874} & \cellcolor[gray]{0.9}{0.841} & \cellcolor[gray]{0.9}{0.745} & \cellcolor[gray]{0.9}{0.936} & \cellcolor[gray]{0.9}{0.867} & \cellcolor[gray]{0.9}{0.100} & \cellcolor[gray]{0.9}{0.875} & \cellcolor[gray]{0.9}{0.831} & \cellcolor[gray]{0.9}{0.686} & \cellcolor[gray]{0.9}{0.976} & \cellcolor[gray]{0.9}{0.862} & 0.104 & 0.879 & 0.837 & 0.698 & 0.976 & 0.866 \\
 & 20 & 0.362 & 0.921 & 0.838 & 0.699 & 0.976 & 0.910 & 0.104 & 0.918 & 0.815 & 0.650 & 0.979 & 0.906 & \cellcolor[gray]{0.9}{0.334} & \cellcolor[gray]{0.9}{0.929} & \cellcolor[gray]{0.9}{0.847} & \cellcolor[gray]{0.9}{0.711} & \cellcolor[gray]{0.9}{0.982} & \cellcolor[gray]{0.9}{0.920} \\ 
\hline

\multirow{4}{*}{\begin{tabular}[c]{@{}c@{}}SWEC-ETHZ\\ iEEG\\ Adult\end{tabular}} & 3 & 0.585 & 0.335 & 0.546 & 0.981 & 0.110 & 0.267 & \cellcolor[gray]{0.9}{0.532} & \cellcolor[gray]{0.9}{0.769} & \cellcolor[gray]{0.9}{0.776} & \cellcolor[gray]{0.9}{0.808} & \cellcolor[gray]{0.9}{0.680} & \cellcolor[gray]{0.9}{0.821} & 0.278 & 0.415 & 0.579 & 0.959 & 0.199 & 0.358 \\
 & 5 & 0.487 & 0.417 & 0.600 & 0.980 & 0.220 & 0.380 & 0.355 & 0.584 & 0.601 & 0.514 & 0.886 & 0.659 & 0.234 & 0.541 & 0.649 & 0.917 & 0.381 & 0.529 \\
 & 10 & 0.231 & 0.717 & 0.763 & 0.871 & 0.655 & 0.731 & 0.131 & 0.455 & 0.472 & 0.311 & 0.992 & 0.509 & 0.196 & 0.751 & 0.768 & 0.841 & 0.695 & 0.766 \\
 & 20 & \cellcolor[gray]{0.9}{0.226} & \cellcolor[gray]{0.9}{0.806} & \cellcolor[gray]{0.9}{0.832} & \cellcolor[gray]{0.9}{0.881} & \cellcolor[gray]{0.9}{0.773} & \cellcolor[gray]{0.9}{0.819} & 0.151 & 0.449 & 0.463 & 0.296 & 0.996 & 0.493 & \cellcolor[gray]{0.9}{0.261} & \cellcolor[gray]{0.9}{0.877} & \cellcolor[gray]{0.9}{0.872} & \cellcolor[gray]{0.9}{0.858} & \cellcolor[gray]{0.9}{0.874} & \cellcolor[gray]{0.9}{0.883} \\ 
\hline

\multirow{4}{*}{\begin{tabular}[c]{@{}c@{}}IEEGP\\ iEEG\\ Adult\end{tabular}} & 3 & \cellcolor[gray]{0.9}{0.289} & \cellcolor[gray]{0.9}{0.753} & \cellcolor[gray]{0.9}{0.753} & \cellcolor[gray]{0.9}{0.727} & \cellcolor[gray]{0.9}{0.884} & \cellcolor[gray]{0.9}{0.787} & 0.308 & 0.636 & 0.535 & 0.952 & 0.118 & 0.542 & 0.376 & 0.720 & 0.720 & 0.769 & 0.474 & 0.760 \\
 & 5 & 0.311 & 0.722 & 0.722 & 0.779 & 0.439 & 0.759 & 0.278 & 0.658 & 0.555 & 0.968 & 0.143 & 0.559 & \cellcolor[gray]{0.9}{0.325} & \cellcolor[gray]{0.9}{0.737} & \cellcolor[gray]{0.9}{0.737} & \cellcolor[gray]{0.9}{0.706} & \cellcolor[gray]{0.9}{0.892} & \cellcolor[gray]{0.9}{0.778} \\
 & 10 & 0.306 & 0.692 & 0.692 & 0.631 & 1 & 0.738 & 0.326 & 0.726 & 0.679 & 0.808 & 0.551 & 0.697 & 0.334 & 0.670 & 0.670 & 0.604 & 1 & 0.712 \\
 & 20 & 0.290 & 0.621 & 0.621 & 0.571 & 0.720 & 0.690 & \cellcolor[gray]{0.9}{0.345} & \cellcolor[gray]{0.9}{0.757} & \cellcolor[gray]{0.9}{0.705} & \cellcolor[gray]{0.9}{0.883} & \cellcolor[gray]{0.9}{0.528} & \cellcolor[gray]{0.9}{0.733} & 0.398 & 0.616 & 0.616 & 0.429 & 0.991 & 0.648 \\ 
\hline

\multirow{4}{*}{\begin{tabular}[c]{@{}c@{}}EIM\\ iEEG\\ Adult\end{tabular}} & 3 & 0.292 & 0.650 & 0.553 & 0.953 & 0.152 & 0.556 & 0.180 & 0.372 & 0.545 & 0.939 & 0.150 & 0.310 & 0.201 & 0.631 & 0.505 & 0.999 & 0.010 & 0.495 \\
 & 5 & 0.279 & 0.568 & 0.459 & 0.893 & 0.025 & 0.468 & 0.280 & 0.577 & 0.670 & 0.904 & 0.436 & 0.575 & 0.203 & 0.654 & 0.538 & 0.989 & 0.087 & 0.543 \\
 & 10 & 0.262 & 0.654 & 0.568 & 0.909 & 0.227 & 0.586 & 0.224 & 0.841 & 0.809 & 0.785 & 0.832 & 0.849 & 0.218 & 0.715 & 0.646 & 0.926 & 0.366 & 0.655 \\
 & 20 & \cellcolor[gray]{0.9}{0.204} & \cellcolor[gray]{0.9}{0.648} & \cellcolor[gray]{0.9}{0.644} & \cellcolor[gray]{0.9}{0.603} & \cellcolor[gray]{0.9}{0.685} & \cellcolor[gray]{0.9}{0.611} & \cellcolor[gray]{0.9}{0.246} & \cellcolor[gray]{0.9}{0.833} & \cellcolor[gray]{0.9}{0.850} & \cellcolor[gray]{0.9}{0.886} & \cellcolor[gray]{0.9}{0.808} & \cellcolor[gray]{0.9}{0.846} & \cellcolor[gray]{0.9}{0.224} & \cellcolor[gray]{0.9}{0.780} & \cellcolor[gray]{0.9}{0.745} & \cellcolor[gray]{0.9}{0.881} & \cellcolor[gray]{0.9}{0.609} & \cellcolor[gray]{0.9}{0.749} \\ 
\hline \hline

\end{tabular}
}}
\end{table*}

\subsection{EEG-level Seizure Detection}
Next, we performed EEG-level seizure detection based on the outputs of the segment-level detector. We summarized the results for the six datasets in Table~\ref{tab:LOSO_EEG_detection}. The EEG-level performance is computed according to MOES, as it is more suitable for clinical practice than existing metrics. We also considered other existing evaluation metrics for comparison in Table~\ref{tab:LOSO_EEG_detection_compare_metrics}.

On the TUH-SZ dataset, the CNN-TRF-BM model leads to the most promising results, followed by the CNN-BM and the CNN-SM model. The CNN-TRF-BM EEG-level seizure detector attained a respectable SEN, PRE, aFPR/h, mFPR/h, and median offset of 0.772, 0.429, 0.425, 0, and -2.125s, respectively. While the aFPR/h is high, the mFPR/h is extremely low. This implies that the aFPR/h is skewed by a small number of EEGs containing an exceptionally huge amount of false detection. While the SEN is similar across all three models, the CNN-TRF-BM model reported the best PRE, which is critical for clinical deployment.

Similarly, we evaluated the EEG-level seizure detectors on the five scalp EEG and iEEG datasets. The CNN models yield high SEN, decent PRE, and low aFPR/h and mFPR/h on the CHB-MIT, SWEC-ETHZ, and EIM datasets. Meanwhile, on the HUH and IEEGP datasets, the model achieves low SEN (0.254 and 0.450, respectively), high PRE (0.841 and 0.917, respectively), and low mFPR/h (0.347 and 0, respectively). The poorer results on the HUH dataset align with our expectations since it is a neonatal dataset. The morphology of neonatal seizures differs vastly from adult seizures. Since the model has been trained on adult scalp EEG, it struggles to detect seizures in neonatal scalp EEGs. Meanwhile, the IEEGP dataset contains some dog iEEGs, which could have different seizure patterns from adult humans. However, we observed that the detection performance is comparable for human and dog EEGs. Hence, the proposed detector can detect some neonate and dog seizures with high PRE, which can be tremendously valuable.

We also determined the detection offset, defined as the average duration between the start time of the seizure and the start time of its corresponding detection (see Table~\ref{tab:LOSO_EEG_detection}), which can be negative. A negative offset does not imply forecasting, as the EEG data is analyzed offline~\cite{cook2013prediction}. Therefore, data from future time intervals are being considered to decide whether an EEG segment is ictal. 

In Table~\ref{tab:LOSO_EEG_detection_compare_metrics}, we compare results for the CNN-TRF-BM model for different evaluation metrics (IMS, OVLP, TAES, and MOES). IMS always leads to the best results, followed by OVLP, MOES, and TAES. The results for MOES are similar to OVLP and IMS, despite MOES having a more stringent condition. This implies that the proposed seizure detector detects most seizures with at least 10s overlap and with 30\% overlap between the seizure and detection. The results for the TAES metric are the lowest: a slight drop in SEN, much lower PRE, and significantly higher aFPR/h and mFPR/h. While there are significant differences across the different performance metrics, the results obtained by MOES are the most appropriate, as it does not lead to overly optimistic or pessimistic results.

\begin{figure*}
\centering
\includegraphics[width=16cm]{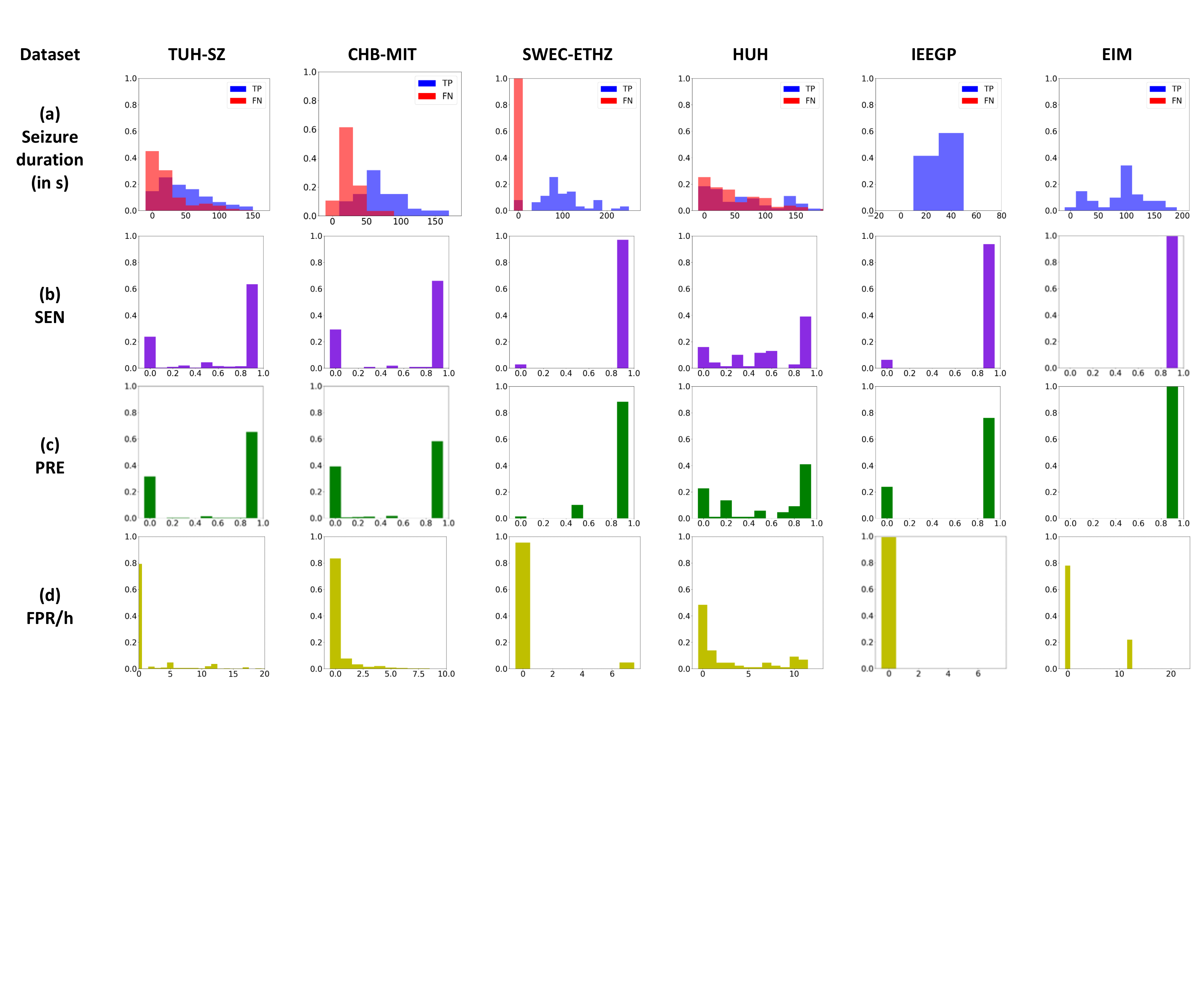}
\vspace*{-3.8cm} 
\caption{EEG-level seizure detection results for the CNN-TRF-BM model across different datasets. (a) normalized histograms of TPs and FNs sorted by seizure duration; (b-d) normalized histograms of the sensitivity (SEN), precision (PRE), and false positive rate per hour (FPR/h) for individual EEGs, respectively.}
\label{fig:Recall_Precision_FPR}
\end{figure*}

Finally, to determine the effectiveness of the CNN-TRF-BM-based EEG-level seizure detector (Figure~\ref{fig:Recall_Precision_FPR}), we plot the normalized histograms of the TP and FN of seizures detected sorted by event duration, together with the normalized histogram of SEN, PRE, and FPR/h computed from individual EEGs across the datasets. From Figure~\ref{fig:Recall_Precision_FPR}(a), it can be seen that it is easier to detect a long seizure than a short event. Figure~\ref{fig:Recall_Precision_FPR}(b) and~\ref{fig:Recall_Precision_FPR}(c) reveal that the SEN and PRE are high for most EEGs, with only a minority of the files having a poor detection rate. Lastly, Figure~\ref{fig:Recall_Precision_FPR}(d) confirms that the system does not make false detections in most EEGs, as mFPR/h is 0. Taken together, these figures suggest that the proposed detector performs well across most EEGs.

In Table~\ref{tab:sensitivity_long_short_seizure_MOES}, we computed the SEN of short ($<$10s) and long ($>$10s) seizures across the six datasets for various window lengths. As the CHB-MIT and IEEGP datasets do not contain annotated short seizures, we could not compute the SEN for those datasets. We observe that $W=20s$ leads to drastic drops in SEN for shorter seizures. On the other hand, a shorter window (3s and 5s) can more reliably capture shorter seizures, at the cost of potentially higher FPR/h.

The proposed seizure detectors, specifically the CNN-TRF-BM-based model, can detect patient-independent seizures at the channel-, segment-, and EEG-level across various scalp EEG and iEEG datasets without retraining. It takes less than 15s computation time to detect seizures in a 30 minutes EEG. Hence, the proposed detector can help automate EEG annotations clinically. However, while the results are appealing for adult human EEG, there is room for improvement for neonatal EEG. One may need to perform additional tuning or retraining to achieve better performance for such cases.

\begin{figurehere}
\centering
\includegraphics[width=8cm]{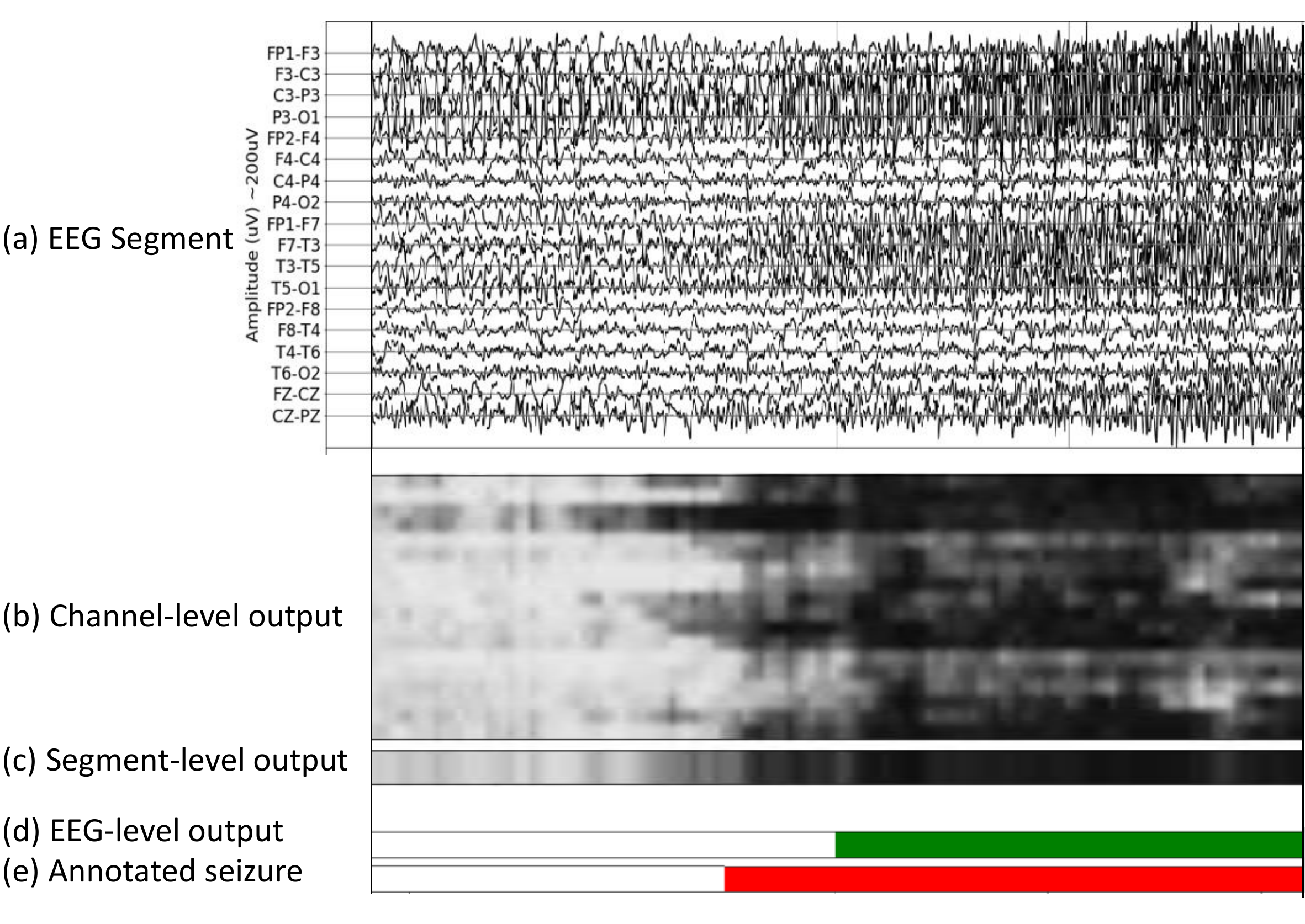}
\caption{Illustration of the outputs of the proposed CNN-TRF-BM seizure detector applied to an EEG. (a)~multi-channel EEG as input; (b)~channel-level predictions; (c)~segment-level predictions; (d)~EEG-level predictions; (e)~annotated seizure.}
\label{fig:system_output_seizure}
\end{figurehere}

\begin{table*}
\centering
\tbl{EEG-level seizure detection results for different models evaluated with MOES across six EEG datasets. \label{tab:LOSO_EEG_detection}}
{
\scalebox{0.63}{
\centering
\begin{tabular}{|c|c|ccccc|ccccc|ccccc|} 
\hline \hline
\multirow{2}{*}{\textbf{Dataset}} & \multirow{2}{*}{\textbf{W}} & \multicolumn{5}{c|}{\textbf{CNN-SM}} & \multicolumn{5}{c|}{\textbf{CNN-BM}} & \multicolumn{5}{c|}{\textbf{CNN-Transformer-BM}} \\ 
\cline{3-17}
 & & \textbf{SEN} & \textbf{PRE} & \textbf{aFPR/h} & \textbf{mFPR/h} & \textbf{Offset} & \textbf{SEN} & \textbf{PRE} & \textbf{aFPR/h} & \textbf{mFPR/h} & \textbf{Offset} & \textbf{SEN} & \textbf{PRE} & \textbf{aFPR/h} & \textbf{mFPR/h} & \textbf{Offset} \\ 
\hline \hline

\multirow{4}{*}{\begin{tabular}[c]{@{}c@{}}TUH-SZ \\Scalp EEG \\Adult\end{tabular}} & 3 & 0.7 & 0.457 & 0.803 & 0 & -4.125 & \cellcolor[gray]{0.9}{0.713} & \cellcolor[gray]{0.9}{0.49} & \cellcolor[gray]{0.9}{0.479} & \cellcolor[gray]{0.9}{0} & \cellcolor[gray]{0.9}{-4.5} & 0.772 & 0.429 & 0.425 & 0 & -2.125 \\
 & 5 & 0.704 & 0.48 & 0.555 & 0 & -4.5 & 0.701 & 0.491 & 0.413 & 0 & -1.5 & 0.653 & 0.476 & 0.411 & 0 & 0.625 \\
 & 10 & \cellcolor[gray]{0.9}{0.719} & \cellcolor[gray]{0.9}{0.495} & \cellcolor[gray]{0.9}{0.466} & \cellcolor[gray]{0.9}{0} & \cellcolor[gray]{0.9}{-1} & 0.701 & 0.512 & 0.237 & 0 & 1.5 & \cellcolor[gray]{0.9}{0.671} & \cellcolor[gray]{0.9}{0.534} & \cellcolor[gray]{0.9}{0.954} & \cellcolor[gray]{0.9}{0} & \cellcolor[gray]{0.9}{0.5} \\
 & 20 & 0.707 & 0.467 & 0.679 & 0 & 6.25 & 0.708 & 0.49 & 0.468 & 0 & 6 & 0.655 & 0.52 & 1.037 & 0 & 2.875 \\ 
\hline

\multirow{4}{*}{\begin{tabular}[c]{@{}c@{}}CHB-MIT \\Scalp EEG \\Pediatric\end{tabular}} & 3 & 0.638 & 0.112 & 1.721 & 1.099 & 0.711 & 0.613 & 0.14 & 0.916 & 0.539 & -2.763 & 0.7 & 0.181 & 1.095 & 0.616 & 0.053 \\
 & 5 & 0.678 & 0.143 & 1.514 & 0.868 & -4.474 & 0.568 & 0.235 & 0.600 & 0.158 & 2.947 & 0.571 & 0.292 & 0.541 & 0.224 & 1.605 \\
 & 10 & 0.734 & 0.254 & 1.041 & 0.618 & 0.158 & \cellcolor[gray]{0.9}{0.704} & \cellcolor[gray]{0.9}{0.411} & \cellcolor[gray]{0.9}{0.291} & \cellcolor[gray]{0.9}{0.026} & \cellcolor[gray]{0.9}{4.737} & \cellcolor[gray]{0.9}{0.678} & \cellcolor[gray]{0.9}{0.377} & \cellcolor[gray]{0.9}{0.421} & \cellcolor[gray]{0.9}{0.118} & \cellcolor[gray]{0.9}{4.684} \\
 & 20 & \cellcolor[gray]{0.9}{0.803} & \cellcolor[gray]{0.9}{0.194} & \cellcolor[gray]{0.9}{1.224} & \cellcolor[gray]{0.9}{0.592} & \cellcolor[gray]{0.9}{6.842} & 0.741 & 0.244 & 0.884 & 0.368 & 5.237 & 0.769 & 0.383 & 0.445 & 0.145 & 1.474 \\ 
\hline

\multirow{4}{*}{\begin{tabular}[c]{@{}c@{}}HUH \\Scalp EEG \\Neonatal\end{tabular}} & 3 & 0.298 & 0.334 & 2.565 & 1.094 & 6 & 0.623 & 0.576 & 2.320 & 2.276 & -3.52 & 0.515 & 0.522 & 2.874 & 2.843 & -2.255 \\
 & 5 & 0.328 & 0.372 & 2.413 & 0.849 & 4.25 & 0.314 & 0.505 & 1.977 & 1.933 & 3.892 & 0.253 & 0.649 & 0.678 & 0.623 & 5.098 \\
 & 10 & 0.254 & 0.397 & 1.671 & 0.181 & 11.5 & \cellcolor[gray]{0.9}{0.214} & \cellcolor[gray]{0.9}{0.807} & \cellcolor[gray]{0.9}{0.334} & \cellcolor[gray]{0.9}{0.303} & \cellcolor[gray]{0.9}{13.52} & \cellcolor[gray]{0.9}{0.227} & \cellcolor[gray]{0.9}{0.818} & \cellcolor[gray]{0.9}{0.253} & \cellcolor[gray]{0.9}{0.223} & \cellcolor[gray]{0.9}{10.853} \\
 & 20 & \cellcolor[gray]{0.9}{0.276} & \cellcolor[gray]{0.9}{0.473} & \cellcolor[gray]{0.9}{1.340} & \cellcolor[gray]{0.9}{0.186} & \cellcolor[gray]{0.9}{14.25} & 0.283 & 0.686 & 0.708 & 0.674 & 16.333 & 0.254 & 0.841 & 0.374 & 0.347 & 15.245 \\ 
\hline

\multirow{4}{*}{\begin{tabular}[c]{@{}c@{}}SWEC-ETHZ \\iEEG \\Adult~\end{tabular}} & 3 & 0.743 & 0.758 & 2.316 & 1.415 & 10.781 & \cellcolor[gray]{0.9}{0.933} & \cellcolor[gray]{0.9}{0.865} & \cellcolor[gray]{0.9}{1.286} & \cellcolor[gray]{0.9}{0.469} & \cellcolor[gray]{0.9}{-2.156} & \cellcolor[gray]{0.9}{0.938} & \cellcolor[gray]{0.9}{0.878} & \cellcolor[gray]{0.9}{0.895} & \cellcolor[gray]{0.9}{0.559} & \cellcolor[gray]{0.9}{7.687} \\
 & 5 & \cellcolor[gray]{0.9}{0.938} & \cellcolor[gray]{0.9}{0.949} & \cellcolor[gray]{0.9}{0.362} & \cellcolor[gray]{0.9}{0} & \cellcolor[gray]{0.9}{3.781} & 0.923 & 0.752 & 2.854 & 2.391 & 0.312 & 0.933 & 0.834 & 1.784 & 1.127 & 4.906 \\
 & 10 & 0.933 & 0.785 & 2.223 & 0.884 & 4.187 & 0.825 & 0.695 & 3.265 & 2.858 & 15.719 & 0.857 & 0.748 & 2.899 & 1.648 & 10.375 \\
 & 20 & 0.878 & 0.711 & 3.897 & 3.601 & 14.937 & 0.911 & 0.744 & 2.764 & 1.259 & 16.531 & 0.849 & 0.727 & 3.010 & 2.205 & 12.5 \\ 
\hline

\multirow{4}{*}{\begin{tabular}[c]{@{}c@{}}IEEGP \\iEEG \\Adult\end{tabular}} & 3 & \cellcolor[gray]{0.9}{0.6} & \cellcolor[gray]{0.9}{0.964} & \cellcolor[gray]{0.9}{0.523} & \cellcolor[gray]{0.9}{0} & \cellcolor[gray]{0.9}{-19} & \cellcolor[gray]{0.9}{0.583} & \cellcolor[gray]{0.9}{0.958} & \cellcolor[gray]{0.9}{0.500} & \cellcolor[gray]{0.9}{0} & \cellcolor[gray]{0.9}{-14.5} & 0.667 & 0.8 & 2.200 & 2 & -19 \\
 & 5 & 0.667 & 0.8 & 2.624 & 2 & -17 & 0.583 & 0.906 & 1.595 & 0 & -17 & \cellcolor[gray]{0.9}{0.617} & \cellcolor[gray]{0.9}{0.944} & \cellcolor[gray]{0.9}{1.120} & \cellcolor[gray]{0.9}{0} & \cellcolor[gray]{0.9}{-17} \\
 & 10 & 0.592 & 0.946 & 0.750 & 0 & -12 & 0.45 & 0.678 & 5.596 & 7 & -12 & 0.5 & 0.753 & 4.423 & 5 & -12 \\
 & 20 & 0.567 & 0.805 & 3.846 & 0 & -2 & 0.542 & 0.906 & 1.500 & 0 & -2 & 0.45 & 0.917 & 0.500 & 0 & -2 \\ 
\hline

\multirow{4}{*}{\begin{tabular}[c]{@{}c@{}}EIM \\iEEG \\Adult\end{tabular}} & 3 & 0.972 & 1 & 0 & 0 & -22.083 & 0.792 & 0.904 & 1.245 & 1.286 & -7 & 1 & 1 & 1.523 & 0 & -32.083 \\
 & 5 & 0.979 & 0.938 & 1.080 & 0 & -30.083 & 1 & 0.972 & 0.484 & 0.452 & -15.417 & \cellcolor[gray]{0.9}{1} & \cellcolor[gray]{0.9}{1} & \cellcolor[gray]{0.9}{0.647} & \cellcolor[gray]{0.9}{0} & \cellcolor[gray]{0.9}{-23.958} \\
 & 10 & \cellcolor[gray]{0.9}{1} & \cellcolor[gray]{0.9}{1} & \cellcolor[gray]{0.9}{0} & \cellcolor[gray]{0.9}{0} & \cellcolor[gray]{0.9}{-23.417} & 0.931 & 0.979 & 0.520 & 0 & 10.542 & 0.931 & 1 & 0.830 & 0 & -1.333 \\
 & 20 & 0.875 & 0.964 & 0.494 & 0.711 & 5.208 & \cellcolor[gray]{0.9}{1} & \cellcolor[gray]{0.9}{1} & \cellcolor[gray]{0.9}{0} & \cellcolor[gray]{0.9}{0} & \cellcolor[gray]{0.9}{-0.792} & 0.951 & 1 & 0.507 & 0 & -3.125 \\ 
\hline \hline

\end{tabular}
}}
\end{table*}

\begin{table*}[htp]
\centering
\tbl{EEG-level seizure detection results by the CNN-TRF-BM-based EEG-level detector evaluated with IMS, OVLP, TAES, and MOES across six EEG datasets. \label{tab:LOSO_EEG_detection_compare_metrics}}
{
\scalebox{0.50}{
\centering
\begin{tabular}{|c|c|ccccc|ccccc|ccccc|ccccc|} 
\hline \hline
\multirow{2}{*}{\textbf{Dataset}} & \multirow{2}{*}{\textbf{W}} & \multicolumn{5}{c|}{\textbf{IMS}} & \multicolumn{5}{c|}{\textbf{OVLP}} & \multicolumn{5}{c|}{\textbf{TAES}} & \multicolumn{5}{c|}{\textbf{MOES}} \\ 
\cline{3-22}
 & & \textbf{SEN} & \textbf{PRE} & \textbf{aFPR/h} & \textbf{mFPR/h} & \textbf{Offset} & \textbf{SEN} & \textbf{PRE} & \textbf{aFPR/h} & \textbf{mFPR/h} & \textbf{Offset} & \textbf{SEN} & \textbf{PRE} & \textbf{aFPR/h} & \textbf{mFPR/h} & \textbf{Offset} & \textbf{SEN} & \textbf{PRE} & \textbf{aFPR/h} & \textbf{mFPR/h} & \textbf{Offset} \\ 
\hline \hline

\multirow{4}{*}{\begin{tabular}[c]{@{}c@{}}TUH-SZ \\scalp EEG \\Adult\end{tabular}} & 3 & \cellcolor[gray]{0.9}{0.797} & \cellcolor[gray]{0.9}{0.437} & \cellcolor[gray]{0.9}{0.412} & \cellcolor[gray]{0.9}{0} & \cellcolor[gray]{0.9}{-7.5} & \cellcolor[gray]{0.9}{0.775} & \cellcolor[gray]{0.9}{0.43} & \cellcolor[gray]{0.9}{0.423} & \cellcolor[gray]{0.9}{0} & \cellcolor[gray]{0.9}{-2} & \cellcolor[gray]{0.9}{0.752} & \cellcolor[gray]{0.9}{0.396} & \cellcolor[gray]{0.9}{0.804} & \cellcolor[gray]{0.9}{0.112} & \cellcolor[gray]{0.9}{-2} & 0.772 & 0.429 & 0.425 & 0 & -2.125 \\
 & 5 & 0.694 & 0.494 & 0.378 & 0 & -4.75 & 0.659 & 0.478 & 0.408 & 0 & 0.5 & 0.652 & 0.42 & 1.001 & 0 & 0.625 & 0.653 & 0.476 & 0.411 & 0 & 0.625 \\
 & 10 & 0.658 & 0.588 & 0.562 & 0 & -2.75 & 0.656 & 0.546 & 0.823 & 0 & 0.5 & 0.66 & 0.435 & 1.840 & 0 & -1.5 & \cellcolor[gray]{0.9}{0.671} & \cellcolor[gray]{0.9}{0.534} & \cellcolor[gray]{0.9}{0.954} & \cellcolor[gray]{0.9}{0} & \cellcolor[gray]{0.9}{0.5} \\
 & 20 & 0.682 & 0.554 & 0.853 & 0 & -2.25 & 0.667 & 0.526 & 1.026 & 0 & 3 & 0.658 & 0.358 & 2.902 & 0 & 3.5 & 0.655 & 0.52 & 1.037 & 0 & 2.875 \\ 
\hline

\multirow{4}{*}{\begin{tabular}[c]{@{}c@{}}CHB-MIT \\scalp EEG \\Pediatric\end{tabular}} & 3 & 0.721 & 0.185 & 1.091 & 0.616 & -5.921 & 0.7 & 0.181 & 1.095 & 0.616 & 0.053 & 0.622 & 0.151 & 1.126 & 0.622 & 0.053 & 0.7 & 0.181 & 1.095 & 0.616 & 0.053 \\
 & 5 & 0.571 & 0.292 & 0.541 & 0.224 & -4.395 & 0.571 & 0.292 & 0.541 & 0.224 & 1.605 & \cellcolor[gray]{0.9}{0.52} & \cellcolor[gray]{0.9}{0.255} & \cellcolor[gray]{0.9}{0.667} & \cellcolor[gray]{0.9}{0.336} & \cellcolor[gray]{0.9}{2.842} & 0.571 & 0.292 & 0.541 & 0.224 & 1.605 \\
 & 10 & 0.672 & 0.434 & 0.244 & 0.053 & 0.789 & \cellcolor[gray]{0.9}{0.668} & \cellcolor[gray]{0.9}{0.402} & \cellcolor[gray]{0.9}{0.359} & \cellcolor[gray]{0.9}{0.118} & \cellcolor[gray]{0.9}{4.895} & 0.567 & 0.179 & 0.780 & 0.414 & 1.237 & \cellcolor[gray]{0.9}{0.678} & \cellcolor[gray]{0.9}{0.377} & \cellcolor[gray]{0.9}{0.421} & \cellcolor[gray]{0.9}{0.118} & \cellcolor[gray]{0.9}{4.684} \\
 & 20 & \cellcolor[gray]{0.9}{0.762} & \cellcolor[gray]{0.9}{0.396} & \cellcolor[gray]{0.9}{0.391} & \cellcolor[gray]{0.9}{0.092} & \cellcolor[gray]{0.9}{-4.368} & 0.769 & 0.383 & 0.445 & 0.145 & 1.474 & 0.597 & 0.14 & 1.272 & 0.623 & 0.921 & 0.769 & 0.383 & 0.445 & 0.145 & 1.474 \\ 
\hline

\multirow{4}{*}{\begin{tabular}[c]{@{}c@{}}SWEC-ETHZ \\iEEG \\Adult\end{tabular}} & 3 & \cellcolor[gray]{0.9}{0.938} & \cellcolor[gray]{0.9}{0.878} & \cellcolor[gray]{0.9}{0.895} & \cellcolor[gray]{0.9}{0.559} & \cellcolor[gray]{0.9}{1.688} & \cellcolor[gray]{0.9}{0.938} & \cellcolor[gray]{0.9}{0.878} & \cellcolor[gray]{0.9}{0.895} & \cellcolor[gray]{0.9}{0.559} & \cellcolor[gray]{0.9}{7.687} & 0.932 & 0.517 & 6.379 & 6.147 & 7.687 & \cellcolor[gray]{0.9}{0.938} & \cellcolor[gray]{0.9}{0.878} & \cellcolor[gray]{0.9}{0.895} & \cellcolor[gray]{0.9}{0.559} & \cellcolor[gray]{0.9}{7.687} \\
 & 5 & 0.933 & 0.853 & 1.620 & 1.127 & -1.094 & 0.933 & 0.84 & 1.729 & 1.127 & 4.906 & 0.913 & 0.523 & 6.863 & 6.259 & 5.187 & 0.933 & 0.834 & 1.784 & 1.127 & 4.906 \\
 & 10 & 0.86 & 0.805 & 2.424 & 1.327 & 11.219 & 0.857 & 0.755 & 2.667 & 1.618 & 10.656 & \cellcolor[gray]{0.9}{0.896} & \cellcolor[gray]{0.9}{0.55} & \cellcolor[gray]{0.9}{6.168} & \cellcolor[gray]{0.9}{5.717} & \cellcolor[gray]{0.9}{6.156} & 0.857 & 0.748 & 2.899 & 1.648 & 10.375 \\
 & 20 & 0.872 & 0.775 & 2.435 & 1.681 & 8.094 & 0.858 & 0.735 & 2.937 & 1.716 & 13.906 & 0.767 & 0.5 & 7.237 & 7.482 & 11.094 & 0.849 & 0.727 & 3.010 & 2.205 & 12.5 \\ 
\hline

\multirow{4}{*}{\begin{tabular}[c]{@{}c@{}}HUH \\scalp EEG \\Neonatal\end{tabular}} & 3 & 0.563 & 0.554 & 2.756 & 2.725 & 0.402 & 0.544 & 0.539 & 2.795 & 2.76 & -1.912 & 0.45 & 0.468 & 3.183 & 3.147 & -1.912 & 0.515 & 0.522 & 2.874 & 2.843 & -2.255 \\
 & 5 & 0.284 & 0.662 & 0.654 & 0.6 & 7.765 & 0.254 & 0.649 & 0.678 & 0.623 & 5.059 & 0.203 & 0.546 & 1.237 & 1.186 & 4.431 & 0.253 & 0.649 & 0.678 & 0.623 & 5.098 \\
 & 10 & \cellcolor[gray]{0.9}{0.203} & \cellcolor[gray]{0.9}{0.918} & \cellcolor[gray]{0.9}{0.053} & \cellcolor[gray]{0.9}{0.041} & \cellcolor[gray]{0.9}{14.392} & \cellcolor[gray]{0.9}{0.215} & \cellcolor[gray]{0.9}{0.822} & \cellcolor[gray]{0.9}{0.228} & \cellcolor[gray]{0.9}{0.203} & \cellcolor[gray]{0.9}{10.186} & \cellcolor[gray]{0.9}{0.18} & \cellcolor[gray]{0.9}{0.564} & \cellcolor[gray]{0.9}{0.628} & \cellcolor[gray]{0.9}{0.593} & \cellcolor[gray]{0.9}{7.843} & \cellcolor[gray]{0.9}{0.227} & \cellcolor[gray]{0.9}{0.818} & \cellcolor[gray]{0.9}{0.253} & \cellcolor[gray]{0.9}{0.223} & \cellcolor[gray]{0.9}{10.853} \\
 & 20 & 0.278 & 0.865 & 0.315 & 0.295 & 17.059 & 0.271 & 0.845 & 0.374 & 0.347 & 16.765 & 0.173 & 0.518 & 1.088 & 1.057 & 18.167 & 0.254 & 0.841 & 0.374 & 0.347 & 15.245 \\ 
\hline

\multirow{4}{*}{\begin{tabular}[c]{@{}c@{}}IEEGP \\iEEG \\Adult\end{tabular}} & 3 & 0.667 & 0.8 & 2.200 & 2 & -37 & 0.667 & 0.8 & 2.200 & 2 & -19 & 0.656 & 0.466 & 15.182 & 15.725 & -19 & 0.667 & 0.8 & 2.200 & 2 & -19 \\
 & 5 & \cellcolor[gray]{0.9}{0.65} & \cellcolor[gray]{0.9}{1} & \cellcolor[gray]{0.9}{0} & \cellcolor[gray]{0.9}{0} & \cellcolor[gray]{0.9}{-35} & \cellcolor[gray]{0.9}{0.625} & \cellcolor[gray]{0.9}{0.958} & \cellcolor[gray]{0.9}{0.750} & \cellcolor[gray]{0.9}{0} & \cellcolor[gray]{0.9}{-17} & 0.592 & 0.547 & 13.431 & 13.571 & -17 & \cellcolor[gray]{0.9}{0.617} & \cellcolor[gray]{0.9}{0.944} & \cellcolor[gray]{0.9}{1.120} & \cellcolor[gray]{0.9}{0} & \cellcolor[gray]{0.9}{-17} \\
 & 10 & 0.55 & 0.822 & 3.096 & 2 & -29 & 0.5 & 0.777 & 3.173 & 0 & -12 & \cellcolor[gray]{0.9}{0.526} & \cellcolor[gray]{0.9}{0.531} & \cellcolor[gray]{0.9}{12.339} & \cellcolor[gray]{0.9}{13.205} & \cellcolor[gray]{0.9}{-12} & 0.5 & 0.753 & 4.423 & 5 & -12 \\
 & 20 & 0.467 & 1 & 0 & 0 & -20 & 0.458 & 0.958 & 0.250 & 0 & -2 & 0.36 & 0.374 & 15.284 & 16.997 & -2 & 0.45 & 0.917 & 0.500 & 0 & -2 \\ 
\hline

\multirow{4}{*}{\begin{tabular}[c]{@{}c@{}}EIM \\iEEG \\Adult\end{tabular}} & 3 & 1 & 1 & 0 & 0 & -41.083 & 1 & 1 & 0 & 0 & -32.083 & 1 & 0.646 & 8.380 & 8.613 & -32.083 & 1 & 1 & 1.523 & 0 & -32.083 \\
 & 5 & \cellcolor[gray]{0.9}{1} & \cellcolor[gray]{0.9}{1} & \cellcolor[gray]{0.9}{0} & \cellcolor[gray]{0.9}{0} & \cellcolor[gray]{0.9}{-32.958} & \cellcolor[gray]{0.9}{1} & \cellcolor[gray]{0.9}{1} & \cellcolor[gray]{0.9}{0} & \cellcolor[gray]{0.9}{0} & \cellcolor[gray]{0.9}{-23.958} & 0.992 & 0.662 & 7.811 & 8.218 & -23.958 & \cellcolor[gray]{0.9}{1} & \cellcolor[gray]{0.9}{1} & \cellcolor[gray]{0.9}{0.647} & \cellcolor[gray]{0.9}{0} & \cellcolor[gray]{0.9}{-23.958} \\
 & 10 & 0.931 & 1 & 0 & 0 & 0.333 & 0.931 & 1 & 0 & 0 & -1.333 & 0.975 & 0.681 & 7.071 & 7.547 & -15.5 & 0.931 & 1 & 0.830 & 0 & -1.333 \\
 & 20 & 0.951 & 1 & 0 & 0 & -12.125 & 0.951 & 1 & 0 & 0 & -3.125 & \cellcolor[gray]{0.9}{0.92} & \cellcolor[gray]{0.9}{0.678} & \cellcolor[gray]{0.9}{6.603} & \cellcolor[gray]{0.9}{6.577} & \cellcolor[gray]{0.9}{-0.208} & 0.951 & 1 & 0.507 & 0 & -3.125 \\ 
\hline \hline

\end{tabular}
}}
\end{table*}

\begin{tablehere}
\tbl{SEN of short ($<$10s) and long ($>$10s) seizures detected by the CNN-TRF-BM-based model across the six datasets according to MOES metric.
\label{tab:sensitivity_long_short_seizure_MOES}}
{
\centering
\scalebox{0.75}{
\begin{tabular}{ccccc} 
\hline \hline
\multirow{2}{*}{\textbf{Dataset}} & \multirow{2}{*}{\textbf{W}} & \multicolumn{3}{c}{\textbf{SEN}} \\ 
\cline{3-5}
 & & \textbf{All Seizure} & \textbf{Short Seizure} & \textbf{Long Seizure} \\ 
\hline \hline

\multirow{4}{*}{\begin{tabular}[c]{@{}c@{}}TUH-SZ \\scalp EEG \\Adult\end{tabular}} & 3 & \cellcolor[gray]{0.9}{0.772} & \cellcolor[gray]{0.9}{0.532} & \cellcolor[gray]{0.9}{0.797} \\
 & 5 & 0.653 & 0.431 & 0.687 \\
 & 10 & 0.671 & 0.343 & 0.72 \\
 & 20 & 0.655 & 0.333 & 0.704 \\ 
\hline

\multirow{4}{*}{\begin{tabular}[c]{@{}c@{}}CHB-MIT \\scalp EEG \\Paediatric\end{tabular}} & 3 & 0.7 & - & 0.75 \\
 & 5 & 0.571 & - & 0.631 \\
 & 10 & 0.678 & - & 0.728 \\
 & 20 & \cellcolor[gray]{0.9}{0.769} & \cellcolor[gray]{0.9}{-} & \cellcolor[gray]{0.9}{0.819} \\ 
\hline

\multirow{4}{*}{\begin{tabular}[c]{@{}c@{}}SWEC-ETHZ \\iEEG \\Adult\end{tabular}} & 3 & \cellcolor[gray]{0.9}{0.938} & \cellcolor[gray]{0.9}{1} & \cellcolor[gray]{0.9}{0.931} \\
 & 5 & 0.933 & 1 & 0.926 \\
 & 10 & 0.857 & 0 & 0.912 \\
 & 20 & 0.849 & 0 & 0.868 \\ 
\hline

\multirow{4}{*}{\begin{tabular}[c]{@{}c@{}}HUH \\scalp EEG \\Neonatal\end{tabular}} & 3 & \cellcolor[gray]{0.9}{0.515} & \cellcolor[gray]{0.9}{0.818} & \cellcolor[gray]{0.9}{0.496} \\
 & 5 & 0.253 & 0.091 & 0.255 \\
 & 10 & 0.227 & 0 & 0.248 \\
 & 20 & 0.254 & 0 & 0.265 \\ 
\hline

\multirow{4}{*}{\begin{tabular}[c]{@{}c@{}}IEEGP \\iEEG \\Adult\end{tabular}} & 3 & \cellcolor[gray]{0.9}{0.667} & \cellcolor[gray]{0.9}{-} & \cellcolor[gray]{0.9}{0.667} \\
 & 5 & 0.617 & - & 0.617 \\
 & 10 & 0.5 & - & 0.5 \\
 & 20 & 0.45 & - & 0.45 \\ 
\hline

\multirow{4}{*}{\begin{tabular}[c]{@{}c@{}}EIM \\iEEG \\Adult\end{tabular}} & 3 & \cellcolor[gray]{0.9}{1} & \cellcolor[gray]{0.9}{1} & \cellcolor[gray]{0.9}{1} \\
 & 5 & 1 & 1 & 1 \\
 & 10 & 0.931 & 0 & 0.955 \\
 & 20 & 0.951 & 0 & 0.955 \\
\hline \hline

\end{tabular}
}}
\end{tablehere}

\section{Discussion}

\subsection{Comparison with Existing Patient-independent Detectors}
To compare the proposed seizure detector to the state-of-the-art is challenging, as there is a lack of standardized evaluation metrics, datasets, or training and testing procedures for the problem of seizure detection. In addition, the datasets considered in the literature vary in terms of patients (age, type, diversity), clinical settings, EEG type, data quantity and quality, and use case (patient-specific vs patient-independent).

It is especially critical to specify the use case, as patient-specific detectors may yield much better performance than a patient-independent detector,  but cannot be readily deployed. Therefore, comparing these two types of detectors is meaningless. Consequently, we consider studies that report patient-independent seizure detection results on the six datasets analyzed in this paper.

\subsubsection{Detection on the TUH-SZ Dataset}
Numerous patient-independent seizure detectors have been evaluated on the TUH-SZ dataset. Roy~\textit{et al.}~utilized different machine learning models and reported a SEN and FPR/h of 0.916 and 137.311~\cite{roy2021evaluation}. Meanwhile, Shah~\textit{et al.}~applied an LTSM to detect seizures at the segment-level and obtained SEN between 0.33-0.37 and FPR/h between 1.24-20.8~\cite{shah2017optimizing}. Ayodele~\textit{et al.}~trained a VGGNet and evaluated it on 24 EEGs, attaining a SEN, FPR/h, and offset of 0.7835, 0.9, and 2.32s, respectively~\cite{ayodele2020supervised}.

Most results reported are not suitable for clinical application; extremely low SEN or high FPR/h. Additionally, most studies did not report the seizure evaluation metrics. When they do, they utilize EBS and OVLP metrics, which fail to represent the requirements of a seizure detector appropriately. In contrast, the proposed CNN-TRF-BM seizure detector achieved superior results calculated with MOES (SEN, PRE, aFPR/h, and mFPR/h of 0.772, 0.429, 0.425, and 0, respectively), which is suitable for clinical applications. However, to the author's knowledge, no existing studies have reported the PRE, although it is an essential metric in clinical practice. Moreover, only a few studies reported the offset.

\subsubsection{Detection on the CHB-MIT Dataset}
In the following, we review the results of the CHB-MIT dataset reported in the literature. Furbass~\textit{et al.}~deployed epileptiform wave sequence (EWS) to classify seizures and obtained a SEN and FPR/h of 0.67 and 0.32, respectively~\cite{furbass2015prospective}. Gómez~\textit{et al.}~applied a CNN and achieved a SEN, SPE, and FPR/h of 0.531, 0.931, and 7.8, respectively~\cite{gomez2020automatic}. Ayodele~\textit{et al.}~employed the CHB-MIT and TUH-SZ dataset and reported a SEN, FPR/h, and offset of 0.7145, 0.76, and 2.32s, respectively~\cite{ayodele2020supervised}. Mansouri~\textit{et al.}~trained their detector on the CHB-MIT (19 patients) and the TUH-SZ (24 patients) dataset and evaluated the detector on the CHB-MIT dataset~\cite{mansouri2019online}. They attained a SEN, SPE, and FPR/h of 0.83, 0.96, and 8, respectively. 

The proposed CNN-TRF-BM model achieves better results on the CHB-MIT dataset, with SEN, PRE, aFPR/h, mFPR/h, and offset of 0.678, 0.377, 0.421, 0.118, and 4.684s, respectively. However, we trained our detector with the TUH-SZ dataset instead of the CHB-MIT dataset. The TUH-SZ dataset contains more seizures (3,055 events) compared to CHB-MIT (185 events), giving the model more data to learn from. This shows that training the detector on a different but larger dataset may help improve the performance.

\subsubsection{Detection on the SWEC-ETHZ Dataset}
No existing seizure detectors had been evaluated on the SWEC-ETHZ dataset in a patient-independent manner. Existing studies only performed patient-specific detection on this dataset~\cite{burrello2019hyperdimensional}. The current study can be the baseline for patient-independent seizure detection on the SWEC-ETHZ dataset.

\subsubsection{Detection on the HUH Dataset}
No seizure detectors have so far been evaluated on the HUH dataset in a patient-independent manner. Existing studies only evaluated patient-specific seizure detection~\cite{o2020neonatal}. The current study is the first to perform patient-independent seizure detection at EEG-level on the HUH dataset. Moreover, we applied a detector trained on adult EEGs to detect seizures in neonatal EEGs and attained promising results. This shows that a detector trained on adult seizures may capture neonatal seizures with a high PRE, despite the substantial age gap. As the model has been trained on adult scalp EEG, it struggles to detect all seizures in neonatal EEGs.

\subsubsection{Detection on the IEEGP Dataset}
Few studies investigated seizure detection on the IEEGP dataset. All studies are on patient-specific seizure detections~\cite{brinkmann2016crowdsourcing}. Similarly, accuracy is a poor metric for an imbalanced dataset. Therefore, the current study can be the baseline for patient-independent seizure detection on the IEEGP dataset.

\subsubsection{Detection on the EIM Dataset}
No earlier studies on seizure detection have been conducted on the EIM dataset. The existing studies aim to predict surgical outcomes~\cite{li2021neural}. The current study is the first to analyze the EIM dataset for patient-independent seizure detection.

\subsection{Commercial Detectors}
Several commercial seizure detectors are available in the market, such as Persyst~\cite{sierra2015seizure}, Encevis~\cite{rommens2018improving}, and BESA~\cite{rommens2018improving}. Earlier studies by Reus~\textit{et al.}~\cite{reus2022automated} and Koren~\textit{et al.}~\cite{koren2021systematic} have compared the performance of Persyst, Encevis, and BESA. We summarized their findings against the performance of the proposed detector in Table~\ref{tab:comparision_commercial_seizure_detector}. Both studies evaluated the commercial detectors on adult scalp EEG datasets; hence, we focus on the TUH-SZ dataset in this section.

The proposed model outperforms the three commercial detectors in the study conducted by Reus~\textit{et al.}~by a significant margin. Meanwhile, the proposed system outperforms Persyst and BESA in the study by Koren~\textit{et al.}, with Encevis reporting similar results to the current study. However, we report MOES, TAES, OVLP, and IMS metric results. In contrast, Reus~\textit{et al.}~and Koren~\textit{et al.}~only reported IMS, which is more lenient as they consider a detection correct as long as the detection is within 30s before the start or after the end of the seizure. Koren~\textit{et al.}~implemented an altered version of IMS, where the margin is increased to 120s. These metrics introduced a significant margin of error, which is inappropriate in clinical practice.

\begin{tablehere}
\tbl{Performance of commercial seizure detectors against the proposed CNN-TRF-BM detector. \label{tab:comparision_commercial_seizure_detector}}
{
\centering
\scalebox{0.55}{
\begin{tabular}{cccccccc} 
\hline\hline
\textbf{Author} & \begin{tabular}[c]{@{}c@{}}\textbf{No of} \\\textbf{Patients}\end{tabular} & \begin{tabular}[c]{@{}c@{}}\textbf{No of} \\\textbf{Seizures}\end{tabular} & \begin{tabular}[c]{@{}c@{}}\textbf{Duration} \\\textbf{(in hours)}\end{tabular} & \textbf{Metrics} & \begin{tabular}[c]{@{}c@{}}\textbf{Seizure} \\\textbf{Detector}\end{tabular} & \textbf{SEN} & \textbf{aFPR/h} \\ 
\hline\hline

\multirow{3}{*}{\begin{tabular}[c]{@{}c@{}}Reus~\textit{et al.}\\\cite{reus2022automated}\end{tabular}} & \multirow{3}{*}{283} & \multirow{3}{*}{249} & \multirow{3}{*}{8771} & \multirow{3}{*}{IMS} & Persyst 14 & 0.558 & 0.071 \\ 
\cline{6-8}
 & & & & & Encevis 1.9.2 & 0.518 & 0.229 \\ 
\cline{6-8}
 & & & & & BESA 2.0 & 0.430 & 0.100 \\ 
\hline

\multirow{3}{*}{\begin{tabular}[c]{@{}c@{}}Koren~\textit{et al.}\\\cite{koren2021systematic}\end{tabular}} & \multirow{3}{*}{81} & \multirow{3}{*}{790} & \multirow{3}{*}{6900} & \multirow{3}{*}{IMS} & Persyst 13 & 0.816 & 0.9 \\ 
\cline{6-8}
 & & & & & Encevis 1.7 & 0.778 & 0.2 \\ 
\cline{6-8}
 & & & & & BESA 2.0 & 0.676 & 0.7 \\ 
\hline\hline

\multirow{3}{*}{\begin{tabular}[c]{@{}c@{}}Current\\study\end{tabular}} & \multirow{3}{*}{\begin{tabular}[c]{@{}c@{}}637\\TUH-SZ\end{tabular}} & \multirow{3}{*}{3055} & \multirow{3}{*}{922} & MOES & CNN-TRF-BM & 0.772 & 0.425 \\ 
\cline{5-8}
 & & & & OVLP & CNN-TRF-BM & 0.775 & 0.423 \\ 
\cline{5-8}
 & & & & IMS & CNN-TRF-BM & 0.797 & 0.412 \\
\hline \hline
\end{tabular}
}}
\end{tablehere}

\subsection{Transformer for Seizure Detection}
We identified two studies that apply transformers for seizure detection~\cite{bhattacharya2021epileptic}. However, these systems did not implement a channel-level detector but headed directly to the segment-level. Thus, the current study is the first to implement a channel-level seizure detector through transformers.

Bhattacharya~\textit{et al.}~utilized a transformer for patient-specific seizure detection on the CHB-MIT and IEEGP dataset~\cite{bhattacharya2021epileptic}. For the CHB-MIT and IEEGP datasets, they attained an average SEN of 0.985 and 0.948, and FPR/h of 0.124 and 0, respectively. While they used transformers, there were significant differences in the study performed by Bhattacharya~\textit{et al.}~as compared to the current study. Firstly, we followed a patient-independent approach while they designed a patient-specific detector. Secondly, The proposed system can detect seizures at the channel-level. In contrast, their systems can only detect seizures at the segment-level. Thirdly, we implemented BM loss while they utilized the SM loss.

\subsection{Training the Detector on the TUH-SZ Dataset Only}
Patient-independent seizure detectors that can be readily deployed without retraining are convenient for clinical practice. To replicate this scenario, we only trained the seizure detectors on the TUH-SZ dataset. Earlier, we showed that the proposed seizure detectors yield good performance on six EEG datasets. However, when testing on an independent dataset, we do not know whether the model trained on the TUH-SZ dataset would yield better performance than a model trained on the test dataset.

To address this, we train and test the seizure detectors on the CHB-MIT dataset as the total length of EEG in that dataset is comparable to those in the TUH-SZ dataset. We report the channel-, segment-, and EEG-level results on the CHB-MIT dataset in Table~\ref{tab:LOSO_detection_CHB_MIT_CS}. The CNN-TRF-BM model yields a SEN, PRE, aFPR/h and mFPR/h of 0.613, 0.088, 0.408, and 0, respectively. In comparison, the model trained on the TUH-SZ dataset yields a SEN, PRE, aFPR/h, and mFPR/h of 0.678, 0.377, 0.421, and 0.118 (see Table~\ref{tab:LOSO_EEG_detection}). While the model trained on the CHB-MIT dataset obtained lower FPR, the model trained on the TUH-SZ dataset attained vastly superior SEN and PRE.

This experiment suggests that training and evaluating a model with the same dataset might not necessarily generate the best results. Here, when tested on the CHB-MIT dataset, the detector trained on the TUH-SZ dataset performed better than the model trained on the CHB-MIT dataset. This is because the TUH-SZ dataset contains more seizures (3,055 events) than the CHB-MIT dataset (185 events), allowing the detector to learn from a more diverse dataset. This experiment also suggests that designing neural network-based patient-independent seizure detectors that generalize well across different datasets is possible.

Using a pretrained seizure detector to perform seizure detection on another dataset is not a new concept~\cite{saab2020weak}. Saab~\textit{et al.}~trained their detector on the TUH-SZ dataset and evaluated it on their private Stanford dataset, and vice versa~\cite{saab2020weak}. However, they did not achieve better results with this approach. In contrast, we showed that obtaining better results on one dataset with the same approach is possible. Furthermore, training detectors on a large variety of seizures from a large number of patients may boost the robustness of the detectors, allowing them to be deployed effectively in clinical practice.

As we will explain in the next section, the models proposed in this paper contain a small number of parameters compared to the models proposed in Saab~\textit{et al.}, therefore, they are less prone to overfitting and are better able to generalize across datasets.

\begin{table*}
\tbl{Channel-, segment-, and EEG-level results trained and tested on the CHB-MIT dataset.
\label{tab:LOSO_detection_CHB_MIT_CS}}
{
\centering
\scalebox{0.6}{
\begin{tabular}{|c|c|c|cccccc|cccccc|ccccc|} 
\hline \hline
\multirow{2}{*}{\textbf{Dataset}} & \multirow{2}{*}{\textbf{Model}} & \multirow{2}{*}{\textbf{W}} & \multicolumn{6}{c|}{\textbf{Channel-level}} & \multicolumn{6}{c|}{\textbf{Segment-level}} & \multicolumn{5}{c|}{\textbf{EEG-level}} \\ 
\cline{4-20}
 & & & \textbf{ECE} & \textbf{ACC} & \textbf{BAC} & \textbf{SEN} & \textbf{SPE} & \textbf{F1} & \textbf{ECE} & \textbf{ACC} & \textbf{BAC} & \textbf{SEN} & \textbf{SPE} & \textbf{F1} & \textbf{SEN} & \textbf{PRE} & \textbf{aFPR/h} & \textbf{mFPR/h} & \textbf{Offset} \\ 
\hline \hline

\multirow{4}{*}{\begin{tabular}[c]{@{}c@{}}CHB-MIT \\Paediatric \\scalp EEG\end{tabular}} & \multirow{4}{*}{\begin{tabular}[c]{@{}c@{}}1D \\CNN-SM\end{tabular}} & 3 & 0.259 & 0.617 & 0.756 & 0.569 & 0.942 & 0.649 & 0.122 & 0.789 & 0.801 & 0.804 & 0.798 & 0.789 & \cellcolor[gray]{0.9}{0.515} & \cellcolor[gray]{0.9}{0.042} & \cellcolor[gray]{0.9}{2.322} & \cellcolor[gray]{0.9}{0.825} & \cellcolor[gray]{0.9}{-16.875} \\
 & & 5 & 0.181 & 0.669 & 0.763 & 0.56 & 0.966 & 0.668 & 0.105 & 0.814 & 0.824 & 0.762 & 0.887 & 0.808 & 0.509 & 0.041 & 2.371 & 1 & -15.750 \\
 & & 10 & \cellcolor[gray]{0.9}{0.126} & \cellcolor[gray]{0.9}{0.786} & \cellcolor[gray]{0.9}{0.816} & \cellcolor[gray]{0.9}{0.743} & \cellcolor[gray]{0.9}{0.889} & \cellcolor[gray]{0.9}{0.79} & \cellcolor[gray]{0.9}{0.118} & \cellcolor[gray]{0.9}{0.874} & \cellcolor[gray]{0.9}{0.841} & \cellcolor[gray]{0.9}{0.745} & \cellcolor[gray]{0.9}{0.936} & \cellcolor[gray]{0.9}{0.867} & 0.509 & 0.037 & 2.588 & 0.875 & -15.750 \\
 & & 20 & 0.129 & 0.777 & 0.78 & 0.592 & 0.969 & 0.758 & 0.362 & 0.921 & 0.838 & 0.699 & 0.976 & 0.910 & 0.509 & 0.040 & 2.433 & 0.875 & -21.750 \\ 
\hline

\multirow{4}{*}{\begin{tabular}[c]{@{}c@{}}CHB-MIT \\Paediatric \\scalp EEG\end{tabular}} & \multirow{4}{*}{\begin{tabular}[c]{@{}c@{}}1D \\CNN-BM\end{tabular}} & 3 & 0.269 & 0.568 & 0.74 & 0.51 & 0.97 & 0.601 & 0.117 & 0.798 & 0.811 & 0.819 & 0.804 & 0.801 & 0.510 & 0.044 & 2.221 & 0.750 & -7.250 \\
 & & 5 & 0.205 & 0.62 & 0.739 & 0.494 & 0.984 & 0.616 & 0.126 & 0.811 & 0.816 & 0.700 & 0.932 & 0.808 & \cellcolor[gray]{0.9}{0.516} & \cellcolor[gray]{0.9}{0.028} & \cellcolor[gray]{0.9}{3.420} & \cellcolor[gray]{0.9}{0.881} & \cellcolor[gray]{0.9}{3.500} \\
 & & 10 & \cellcolor[gray]{0.9}{0.137} & \cellcolor[gray]{0.9}{0.724} & \cellcolor[gray]{0.9}{0.782} & \cellcolor[gray]{0.9}{0.635} & \cellcolor[gray]{0.9}{0.928} & \cellcolor[gray]{0.9}{0.733} & \cellcolor[gray]{0.9}{0.100} & \cellcolor[gray]{0.9}{0.875} & \cellcolor[gray]{0.9}{0.831} & \cellcolor[gray]{0.9}{0.686} & \cellcolor[gray]{0.9}{0.976} & \cellcolor[gray]{0.9}{0.862} & 0.503 & 0.033 & 2.894 & 1 & -7.000 \\
 & & 20 & 0.141 & 0.777 & 0.782 & 0.606 & 0.959 & 0.765 & 0.104 & 0.918 & 0.815 & 0.650 & 0.979 & 0.906 & 0.515 & 0.033 & 2.947 & 1 & -22.500 \\ 
\hline

\multirow{4}{*}{\begin{tabular}[c]{@{}c@{}}CHB-MIT \\Paediatric \\scalp EEG\end{tabular}} & \multirow{4}{*}{\begin{tabular}[c]{@{}c@{}}1D \\CNN-TRF-BM\end{tabular}} & 3 & 0.25 & 0.582 & 0.747 & 0.528 & 0.966 & 0.617 & \cellcolor[gray]{0.9}{0.258} & \cellcolor[gray]{0.9}{0.833} & \cellcolor[gray]{0.9}{0.847} & \cellcolor[gray]{0.9}{0.808} & \cellcolor[gray]{0.9}{0.886} & \cellcolor[gray]{0.9}{0.837} & 0.577 & 0.214 & 0.101 & 0 & 9.750 \\
 & & 5 & \cellcolor[gray]{0.9}{0.095} & \cellcolor[gray]{0.9}{0.742} & \cellcolor[gray]{0.9}{0.808} & \cellcolor[gray]{0.9}{0.666} & \cellcolor[gray]{0.9}{0.95} & \cellcolor[gray]{0.9}{0.755} & 0.256 & 0.822 & 0.824 & 0.715 & 0.932 & 0.819 & \cellcolor[gray]{0.9}{0.613} & \cellcolor[gray]{0.9}{0.088} & \cellcolor[gray]{0.9}{0.408} & \cellcolor[gray]{0.9}{0} & \cellcolor[gray]{0.9}{3.625} \\
 & & 10 & 0.205 & 0.663 & 0.748 & 0.515 & 0.981 & 0.649 & 0.104 & 0.879 & 0.837 & 0.698 & 0.976 & 0.866 & 0.515 & 0.367 & 0.108 & 0 & -3.375 \\
 & & 20 & 0.153 & 0.755 & 0.756 & 0.534 & 0.978 & 0.733 & 0.334 & 0.929 & 0.847 & 0.711 & 0.982 & 0.920 & 0.568 & 0.478 & 0.041 & 0 & 1 \\
\hline \hline
\end{tabular}
}}
\end{table*}

\subsection{Complexity of Seizure Detectors}
Most seizure detectors proposed in the literature do not perform channel-level detection and proceed to segment-level detection directly. The main innovation in those studies lies in improving the deep neural networks used for segment-level classification. These deep neural networks typically contain numerous layers (often 10+) and millions of parameters, which require substantial computational power for training. Moreover, such networks tend to overfit specific datasets, leading to poor generalization. We explore whether deeper models lead to better seizure detection performance.

In Table~\ref{tab:SZ_complexity_literature}, we list different deep learning systems and provide information about their design and seizure detection performance. These neural networks for seizure detectors contain many layers, ranging between 2 and 709, and contain 7,600 to 138 million parameters. The inputs to those models also vary significantly, ranging from 5,888 to 228,000 input data points. In contrast, the three proposed seizure detectors only require between 384 to 2,560 input data points for window lengths varying from 5s to 20s. Moreover, the models contain 7 to 15 layers, with 0.16 to 3.5 million parameters for the CNN and CNN-TRF models. The input size, number of layers, and parameters for the proposed models are much smaller than for most of the existing models listed in Table~\ref{tab:SZ_complexity_literature}.

Next, we examined the correlation between model size and performance. The proposed seizure detector models reported higher SEN and lower FPR/h than most models with more parameters and layers. The AUC, ACC, BAC, and F1 were comparable, while the SPE was poorer in our model. However, SPE is only computed in segment-level classification, which is not an EEG-level detection metric. Moreover, the proposed models obtained better AUPRC, SEN, and FPR/h than most existing models with fewer parameters and layers.

Overall, the proposed models outshine models with vastly more parameters, which suggests that designing ever-bigger neural networks for seizure detection may not be a fruitful avenue for research. Instead, alternative pipelines with substantially fewer parameters may perform comparably to the state-of-the-art or even better. In this study, we demonstrated that by first detecting seizures at individual channels, one could vastly reduce the number of parameters while achieving the same or increased level of performance.

\begin{table*}[htb]
\centering
\tbl{Deep learning models in the literature in terms of complexity and performance. \label{tab:SZ_complexity_literature}}
{
\scalebox{0.6}{
\begin{tabular}{cccccccccccccc} 

\hline \hline

\textbf{Author} & \textbf{Model} & \textbf{Layers} & \begin{tabular}[c]{@{}c@{}}\textbf{Parameters}\\\textbf{(in millions)}\end{tabular} & \begin{tabular}[c]{@{}c@{}}\textbf{Input}\\\textbf{Size}\end{tabular} & \textbf{AUC} & \textbf{AUPRC} & \textbf{ACC} & \textbf{BAC} & \textbf{SEN} & \textbf{SPE} & \textbf{PRE} & \textbf{F1} & \textbf{FPR/h} \\ 
\hline \hline

Asif~\textit{et al.}~\cite{asif2020seizurenet} & SeizureNet & 133 & 45.94 & 150,528 & - & - & - & - & - & - & - & 0.896 & - \\ 
\hline

\multirow{10}{*}{Raghu~\textit{et al.}~\cite{raghu2020eeg}} & AlexNet & 25 & 62 & 51,529 & - & - & 0.768 & - & - & - & - & - & - \\
 & VGG16 & 41 & 138 & 50,176 & - & - & 0.833 & - & - & - & - & - & - \\
 & VGG19 & 47 & 138 & 50,176 & - & - & 0.818 & - & - & - & - & - & - \\
 & SqueezeNet & 68 & 1.2 & 51,529 & - & - & 0.851 & - & - & - & - & - & - \\
 & GoogleNet & 144 & 7 & 50,176 & - & - & 0.745 & - & - & - & - & - & - \\
 & Inceptionv3 & 316 & 24 & 89,401 & - & - & 0.883 & - & - & - & - & - & - \\
 & DenseNet201 & 709 & 20 & 50,176 & - & - & 0.851 & - & - & - & - & - & - \\
 & ResNet18 & 72 & 11 & 50,176 & - & - & 0.862 & - & - & - & - & - & - \\
 & ResNet50 & 177 & 23 & 50,176 & - & - & 0.862 & - & - & - & - & - & - \\
 & ResNet101 & 347 & 29.4 & 50,176 & - & - & 0.863 & - & - & - & - & - & - \\ 
\hline

Covert~\textit{et al.}~\cite{covert2019temporal} & TGCN & 30 & 5.5 & 415,107 & 0.926 & - & - & 0.809 & 0.648 & 0.970 & - & - & - \\
\hline

%Hossain~\textit{et al.}~\cite{hossain2019applying} & CNN & 4 & 0.04 & 11,500 & - & - & - & 0.908 & 0.900 & 0.917 & - & - & - \\ 
\hline

Yuan~\textit{et al.}~\cite{yuan2018multi} & CNN & 4 & 0.04 & 17,664 & 0.957 & 0.906 & 0.944 & - & - & - & - & 0.853 & - \\ 
\hline

Zhou~\textit{et al.}~\cite{zhou2018epileptic} & CNN & 3 & 0.4 & 5,888 & - & - & 0.595 & 0.595 & 0.618 & 0.572 & - & - & - \\ 
\hline

Saab~\textit{et al.}~\cite{saab2020weak} & ChronoNet & 10 & 12.7 & 45,600 & 0.930 & - & - & - & - & - & - & 0.770 & 0.100 \\ 
\hline

Emami~\textit{et al.}~\cite{emami2019seizure} & VGG16 & 41 & 138 & 50,176 & - & - & - & - & 0.740 & - & - & - & 0.200 \\ 
\hline

Ansari~\textit{et al.}~\cite{ansari2019neonatal} & CNN & 23 & 0.0076 & 54,000 & 0.830 & - & - & - & 0.770 & - & - & - & 0.900 \\ 
\hline

%Khalkhali~\textit{et al.}~\cite{khalkhali2021low} & ResNet & 72 & 11 & 65,536 & - & - & - & - & 0.421 & - & - & - & 0.241 \\ 
\hline

%Shah~\textit{et al.}~\cite{shah2021improved} & CNN-LSTM & 7 & 0.5 & 77,440 & - & - & - & 0.638 & 0.308 & 0.969 & - & - & 0.281 \\
\hline

Gomez~\textit{et al.}~\cite{gomez2020automatic} & CNN & 12 & 0.314 & 21,504 & - & 0.440 & 0.929 & 0.731 & 0.531 & 0.931 & 0.514 & 0.461 & 7.800 \\ 
\hline

%Yang~\textit{et al.}~\cite{yang2020two} & CNN-LTSM & 6 & 0.4 & 34,038 & - & - & - & - & 0.020 & - & - & - & 0.007 \\ 
\hline

%Chatzichristos~\textit{et al.}~\cite{chatzichristos2020epileptic} & U-Net & 16 & 5.88 & 76,000 & - & - & - & - & 0.124 & - & - & - & 0.060 \\ 
\hline \hline

\multirow{4}{*}{Current study} & CNN & 7 & 0.16 & 384 & - & - & - & - & 0.713 & - & 0.490 & 0.581 & 0 \\
 & CNN & 7 & 0.26 & 640 & - & - & - & - & 0.701 & - & 0.491 & 0.578 & 0 \\
 & CNN & 7 & 0.52 & 1,280 & - & - & - & - & 0.701 & - & 0.512 & 0.592 & 0 \\
 & CNN & 7 & 1 & 2,560 & - & - & - & - & 0.708 & - & 0.490 & 0.579 & 0 \\ 
\hline

\multirow{4}{*}{Current study} & CNN-TRF & 15 & 2.3 & 384 & - & - & - & - & 0.772 & - & 0.429 & 0.552 & 0 \\
 & CNN-TRF & 15 & 2.5 & 640 & - & - & - & - & 0.653 & - & 0.476 & 0.551 & 0 \\
 & CNN-TRF & 15 & 2.8 & 1,280 & - & - & - & - & 0.671 & - & 0.534 & 0.595 & 0 \\
 & CNN-TRF & 15 & 3.5 & 2,560 & - & - & - & - & 0.655 & - & 0.520 & 0.580 & 0 \\
\hline \hline

\end{tabular}
}}
\end{table*}

\begin{figure*}[htp]
\begin{minipage}[b]{0.49\linewidth}
  \centering
  \centerline{\includegraphics[width=7cm]{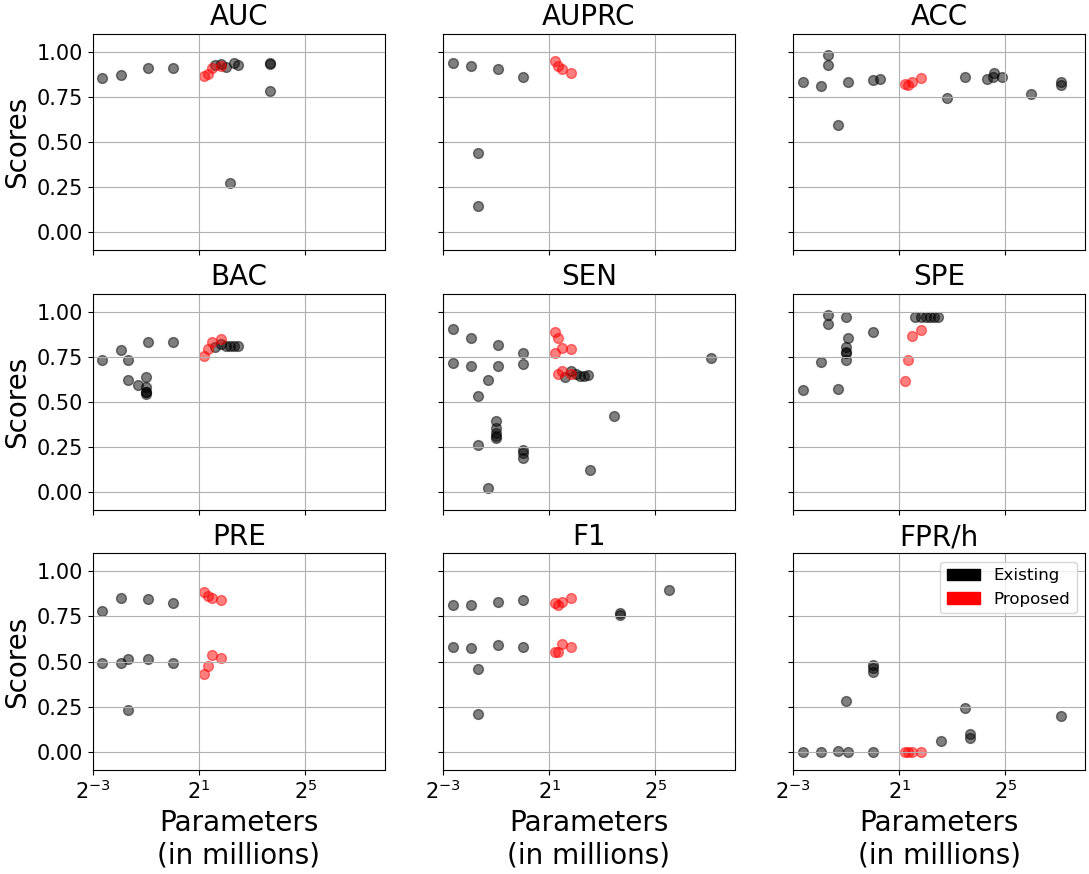}}
  \centerline{(a) Performance vs. number of parameters.}\medskip
\end{minipage}
\hfill
\begin{minipage}[b]{0.49\linewidth}
  \centering
  \centerline{\includegraphics[width=7.5cm]{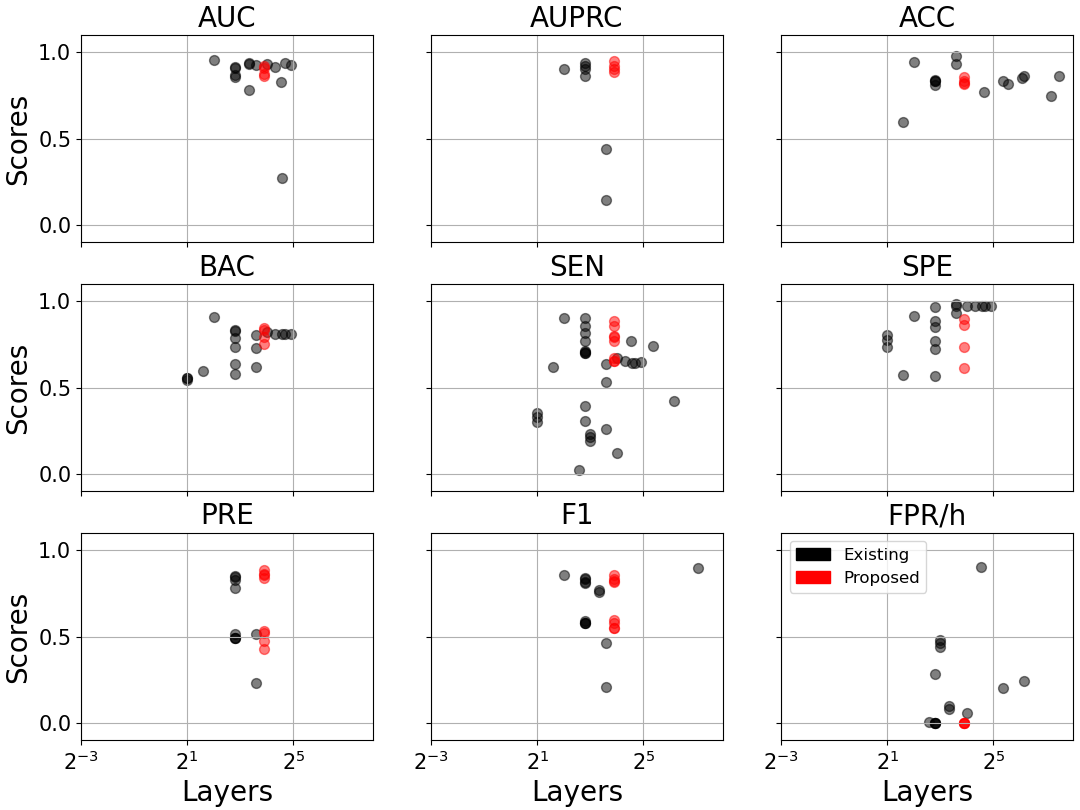}}
  \centerline{(b) Performance vs. number of layers.}\medskip
\end{minipage}

\vspace{-0.2cm}
\caption{Performance of various seizure detectors as a function of (a) parameters (in millions) and (b) layers in the deep learning model. Each red indicates a model (any of the three models) deployed in the current study on the TUH-SZ dataset. At the same time, each black dot denotes a model in the literature~\cite{asif2020seizurenet, raghu2020eeg, covert2019temporal, yuan2018multi, zhou2018epileptic, saab2020weak, emami2019seizure, ansari2019neonatal, gomez2020automatic}. The x-axis in all plots is in the logarithmic scale of base 2.}
\label{fig:complexity_plot}
\end{figure*}

\subsection{Benefits of Channel-level Detection}

In this study, we proposed to detect seizures starting at channel-level. However, many seizure detectors in the literature detect seizures directly from segment-level. This approach usually requires a fixed number of channels, an important limiting factor for clinical practice as the number of channels may vary. Moreover, this approach may be strongly overfitted to a particular EEG electrode configuration and dataset. Here, we refer to seizure detector pipelines that detect seizures starting from channel-level as 1D models, while detectors that detect seizures starting from segment-level as 2D models.

To further evaluate the benefits of the three proposed 1D seizure detectors, we designed two 2D seizure detectors that directly perform segment-level classification from the EEG signals. Those 2D models are identical to the 1D ones, except that the 1D convolutional filters are replaced with 2D filters. We optimized the 2D CNNs on the TUH SZ dataset with the SM and BM loss, leading to the two different 2D seizure detectors. As all the scalp EEGs in the TUH-SZ dataset contain 20 common channels, we fixed the number of channels to 20. Hence, the input of the 2D segment-level CNN models has dimension $(W*128 \times 20)$, where $W \in \{3, 5, 10, 20 \}$ is the window length. Finally, we combined the 2D segment-level detectors with the same EEG-level pipeline to compute the EEG-level detections.

We trained and evaluated the 2D models on the TUH-SZ dataset for the segment- and EEG-level seizure detection. Next, we deployed the 2D models trained on the TUH-SZ dataset to detect seizures in the CHB-MIT dataset; in this way, we evaluate the model's generalizability. We only consider the CHB-MIT dataset for the assessment, as it is the only dataset with EEGs with the same 20 common channels. Finally, we trained and evaluated the 2D models on the CHB-MIT dataset, and compared them to the models trained on the TUH-SZ dataset.

We display the segment- and EEG-level results for the TUH-SZ and CHB-MIT datasets in Table~\ref{tab:SZ_2D_TUH}, where the EEG-level results are computed by the MOES metric. When trained and evaluated on the TUH-SZ dataset, the 2D models attain much weaker results for both segment- and EEG-level classification than the 1D models (see Table~\ref{tab:LOSO_EEG_detection} for comparison). Those models also perform poorly on the CHB-MIT dataset, leading to substantially lower SEN and PRE scores than the 1D models. Moreover, when the 2D models were trained and evaluated on the CHB-MIT dataset, we obtained the worst results thus far, with PRE lower than 5\% for all cases. These numerical results are in line with many 2D models reported in the literature~\cite{zhou2018epileptic}. Overall, 2D models underperform compared to the 1D models by a considerable margin. In conclusion, the channel-level detector appears vital for superior generalization performance.

\begin{table*}[htb]
\centering
\tbl{Results of 2D seizure detectors on the TUH-SZ and CHB-MIT dataset. 
\label{tab:SZ_2D_TUH}}
{
\scalebox{0.6}{
\begin{tabular}{|c|c|c|c|cccccc|cccccc|} 
\hline \hline
\multirow{2}{*}{\begin{tabular}[c]{@{}c@{}}\textbf{Testing}\\\textbf{Dataset}\end{tabular}} & \multirow{2}{*}{\begin{tabular}[c]{@{}c@{}}\textbf{Training}\\\textbf{Dataset}\end{tabular}} & \multirow{2}{*}{\textbf{Model}} & \multirow{2}{*}{\textbf{W}} & \multicolumn{6}{c|}{\textbf{Segment-level}} & \multicolumn{6}{c|}{\textbf{EEG-level}} \\ 
\cline{5-16}
 & & & & \textbf{ECE} & \textbf{ACC} & \textbf{BAC} & \textbf{SEN} & \textbf{SPE} & \textbf{F1} & \textbf{F1} & \textbf{SEN} & \textbf{PRE} & \textbf{aFPR/h} & \textbf{mFPR/h} & \textbf{Offset} \\ 
\hline \hline

\multirow{4}{*}{\begin{tabular}[c]{@{}c@{}}TUH-SZ\\ EEG\\ Adult\end{tabular}} & \multirow{4}{*}{TUH-SZ} & \multirow{4}{*}{\begin{tabular}[c]{@{}c@{}}2D \\CNN-SM\end{tabular}} & 3 & \cellcolor[gray]{0.9}{0.106} & \cellcolor[gray]{0.9}{0.769} & \cellcolor[gray]{0.9}{0.772} & \cellcolor[gray]{0.9}{0.717} & \cellcolor[gray]{0.9}{0.827} & \cellcolor[gray]{0.9}{0.770} & \cellcolor[gray]{0.9}{0.544} & \cellcolor[gray]{0.9}{0.659} & \cellcolor[gray]{0.9}{0.463} & \cellcolor[gray]{0.9}{2.555} & \cellcolor[gray]{0.9}{0} & \cellcolor[gray]{0.9}{9.125} \\
 & & & 5 & 0.119 & 0.791 & 0.769 & 0.672 & 0.866 & 0.788 & 0.538 & 0.674 & 0.448 & 2.764 & 0 & 7.000 \\
 & & & 10 & 0.149 & 0.849 & 0.751 & 0.566 & 0.937 & 0.842 & 0.530 & 0.656 & 0.444 & 3.020 & 0 & 7.500 \\
 & & & 20 & 0.160 & 0.859 & 0.734 & 0.534 & 0.933 & 0.854 & 0.510 & 0.521 & 0.499 & 1.741 & 0 & 3.250 \\ 
\hline

\multirow{4}{*}{\begin{tabular}[c]{@{}c@{}}TUH-SZ\\ EEG\\ Adult\end{tabular}} & \multirow{4}{*}{TUH-SZ} & \multirow{4}{*}{\begin{tabular}[c]{@{}c@{}}2D \\CNN-BM\end{tabular}} & 3 & \cellcolor[gray]{0.9}{0.106} & \cellcolor[gray]{0.9}{0.801} & \cellcolor[gray]{0.9}{0.805} & \cellcolor[gray]{0.9}{0.753} & \cellcolor[gray]{0.9}{0.857} & \cellcolor[gray]{0.9}{0.801} & 0.574 & 0.659 & 0.509 & 2.095 & 0 & -5.250 \\
 & & & 5 & 0.119 & 0.816 & 0.791 & 0.678 & 0.904 & 0.812 & \cellcolor[gray]{0.9}{0.584} & \cellcolor[gray]{0.9}{0.668} & \cellcolor[gray]{0.9}{0.519} & \cellcolor[gray]{0.9}{2.068} & \cellcolor[gray]{0.9}{0} & \cellcolor[gray]{0.9}{-7.875} \\
 & & & 10 & 0.149 & 0.857 & 0.782 & 0.635 & 0.929 & 0.854 & 0.559 & 0.658 & 0.486 & 2.402 & 0 & -3.750 \\
 & & & 20 & 0.160 & 0.868 & 0.711 & 0.468 & 0.955 & 0.856 & 0.528 & 0.506 & 0.553 & 1.499 & 0 & 5.250 \\ 
\hline \hline

\multirow{4}{*}{\begin{tabular}[c]{@{}c@{}}CHB-MIT \\Paediatric \\scalp EEG\end{tabular}} & \multirow{4}{*}{CHB-MIT} & \multirow{4}{*}{\begin{tabular}[c]{@{}c@{}}2D \\CNN-SM\end{tabular}} & 3 & 0.173 & 0.717 & 0.739 & 0.681 & 0.797 & 0.71 & \cellcolor[gray]{0.9}{0.078} & \cellcolor[gray]{0.9}{0.520} & \cellcolor[gray]{0.9}{0.042} & \cellcolor[gray]{0.9}{2.221} & \cellcolor[gray]{0.9}{0.134} & \cellcolor[gray]{0.9}{-16.875} \\
 & & & 5 & 0.143 & 0.732 & 0.748 & 0.638 & 0.858 & 0.714 & 0.079 & 0.503 & 0.043 & 3.420 & 0.520 & -17.250 \\
 & & & 10 & 0.122 & 0.782 & 0.755 & 0.639 & 0.871 & 0.765 & 0.076 & 0.503 & 0.041 & 2.894 & 0.201 & -16.000 \\
 & & & 20 & \cellcolor[gray]{0.9}{0.168} & \cellcolor[gray]{0.9}{0.872} & \cellcolor[gray]{0.9}{0.786} & \cellcolor[gray]{0.9}{0.636} & \cellcolor[gray]{0.9}{0.936} & \cellcolor[gray]{0.9}{0.861} & 0.086 & 0.509 & 0.047 & 2.947 & 0.418 & -16.250 \\ 
\hline

\multirow{4}{*}{\begin{tabular}[c]{@{}c@{}}CHB-MIT \\Paediatric \\scalp EEG\end{tabular}} & \multirow{4}{*}{CHB-MIT} & \multirow{4}{*}{\begin{tabular}[c]{@{}c@{}}2D \\CNN-BM\end{tabular}} & 3 & 0.165 & 0.716 & 0.732 & 0.687 & 0.776 & 0.707 & \cellcolor[gray]{0.9}{0.082} & \cellcolor[gray]{0.9}{0.510} & \cellcolor[gray]{0.9}{0.044} & \cellcolor[gray]{0.9}{2.221} & \cellcolor[gray]{0.9}{0.750} & \cellcolor[gray]{0.9}{-7.250} \\
 & & & 5 & 0.145 & 0.733 & 0.748 & 0.655 & 0.841 & 0.716 & 0.061 & 0.505 & 0.032 & 2.951 & 0.769 & -4.750 \\
 & & & 10 & 0.112 & 0.779 & 0.75 & 0.633 & 0.868 & 0.759 & 0.063 & 0.508 & 0.034 & 2.892 & 1.023 & -7.375 \\
 & & & 20 & \cellcolor[gray]{0.9}{0.16} & \cellcolor[gray]{0.9}{0.868} & \cellcolor[gray]{0.9}{0.783} & \cellcolor[gray]{0.9}{0.635} & \cellcolor[gray]{0.9}{0.932} & \cellcolor[gray]{0.9}{0.858} & 0.067 & 0.510 & 0.036 & 2.692 & 0.848 & -22.250 \\ 
\hline \hline

\multirow{4}{*}{\begin{tabular}[c]{@{}c@{}}CHB-MIT \\Paediatric \\scalp EEG\end{tabular}} & \multirow{4}{*}{TUH-SZ} & \multirow{4}{*}{\begin{tabular}[c]{@{}c@{}}2D \\CNN-SM\end{tabular}} & 3 & 0.233 & 0.584 & 0.662 & 0.365 & 0.959 & 0.547 & 0.303 & 0.439 & 0.231 & 0.391 & 0.040 & -2.737 \\
 & & & 5 & 0.159 & 0.677 & 0.680 & 0.429 & 0.931 & 0.646 & 0.292 & 0.626 & 0.190 & 1.372 & 0.997 & -4.658 \\
 & & & 10 & \cellcolor[gray]{0.9}{0.148} & \cellcolor[gray]{0.9}{0.744} & \cellcolor[gray]{0.9}{0.714} & \cellcolor[gray]{0.9}{0.290} & \cellcolor[gray]{0.9}{0.981} & \cellcolor[gray]{0.9}{0.690} & 0.383 & 0.536 & 0.298 & 0.821 & 0.421 & -2.447 \\
 & & & 20 & 0.083 & 0.843 & 0.658 & 0.265 & 0.991 & 0.798 & \cellcolor[gray]{0.9}{0.370} & \cellcolor[gray]{0.9}{0.376} & \cellcolor[gray]{0.9}{0.365} & \cellcolor[gray]{0.9}{0.113} & \cellcolor[gray]{0.9}{0} & \cellcolor[gray]{0.9}{-5.184} \\ 
\hline

\multirow{4}{*}{\begin{tabular}[c]{@{}c@{}}CHB-MIT \\Paediatric \\scalp EEG\end{tabular}} & \multirow{4}{*}{TUH-SZ} & \multirow{4}{*}{\begin{tabular}[c]{@{}c@{}}2D \\CNN-BM\end{tabular}} & 3 & 0.357 & 0.515 & 0.613 & 0.239 & 0.987 & 0.451 & \cellcolor[gray]{0.9}{0.426} & \cellcolor[gray]{0.9}{0.368} & \cellcolor[gray]{0.9}{0.505} & \cellcolor[gray]{0.9}{0.129} & \cellcolor[gray]{0.9}{0} & \cellcolor[gray]{0.9}{1.526} \\
 & & & 5 & 0.383 & 0.524 & 0.541 & 0.084 & 0.998 & 0.416 & 0.115 & 0.078 & 0.218 & 0.072 & 0 & -0.395 \\
 & & & 10 & 0.156 & 0.750 & 0.648 & 0.307 & 0.989 & 0.696 & 0.441 & 0.524 & 0.380 & 0.396 & 0.050 & -0.737 \\
 & & & 20 & \cellcolor[gray]{0.9}{0.068} & \cellcolor[gray]{0.9}{0.853} & \cellcolor[gray]{0.9}{0.691} & \cellcolor[gray]{0.9}{0.341} & \cellcolor[gray]{0.9}{0.987} & \cellcolor[gray]{0.9}{0.817} & 0.474 & 0.461 & 0.488 & 0.183 & 0.026 & -8.526 \\
\hline \hline
\end{tabular}
}}
\end{table*}

\section{Conclusions and Future work}

This study proposed patient-independent seizure detectors that identify seizures on three EEG scales: channel-, segment- and EEG-level. Firstly, the channel-level detectors detect seizures in single-channel segments through a CNN-based deep learning model. Next, we perform segment-level detection based on statistical features extracted from the channel-level outputs based on different scalp regions. At last, we apply post-processing filters to the segment-level outputs to determine any detected seizures' start and end times. 

We trained and tested the proposed detectors on the TUH-SZ scalp EEG dataset before evaluating the pretrained detectors on five independent scalp EEG and iEEG datasets. Also, we introduced MOES to address some shortcomings of the existing EEG-level seizure detection metrics. To the author's knowledge, this study is one of the first to incorporate a channel-level detector within the seizure detection system~\cite{lu2018classification}. Moreover, we implemented a pipeline that can detect EEG seizures with any number of electrodes. Furthermore, we demonstrated that a channel-level detector is essential for reliable seizure detection and boosting the generalization performance. Finally, the proposed seizure detector is computationally efficient, with a computation time of less than 15s for a 30 minutes EEG. Hence, the detector may help accelerate and improve EEG annotation in clinical practice.

However, as the seizure detector is based on deep learning, it is nearly impossible to identify the exact features or motifs that contribute significantly to the discrimination process. In future work, we can perform feature extraction before deploying the deep learning models. For instance, we can decompose the time series into different frequency bands. This way, we may understand the contribution and significance of each frequency component of the EEG signals.

Additionally, we will address the problem of detecting artifacts before seizure detection~\cite{peh2022transformer}. The artifact detector will be designed to reduce FPR/h and improve the PRE of the seizure detector. Consequently, it can reject artifacts without eliminating important cerebral signals, such as slow waves, sharp waves, and seizures in EEGs. Lastly, we will look into newer and more powerful supervised classification algorithms such as finite element machine and dynamic ensemble algorithm~\cite{pereira2020fema, alam2020dynamic}.

\section*{Conflicts of Interest} 
\noindent The authors have no disclosures to report.

%\section*{Acknowledgments} 
%\noindent The NUH and NNI datasets were collected under the supervision of Dr. Rahul Rathakrishnan and Dr. Yee-Leng Tan, respectively, supported by the National Health Innovation Centre (NHIC) grant (NHIC-I2D-1608138).

%\section*{Appendix}
%\subsection*{A.}
%\subsection*{B.}

\bibliographystyle{ws-ijns}

\begin{thebibliography}{10}

\bibitem{jirsa2014nature}
V.~K. Jirsa, W.~C. Stacey, P.~P. Quilichini, A.~I. Ivanov and C.~Bernard, On
  the nature of seizure dynamics, {\em Brain} {\bf 137}(8)  (2014)  2210--2230.

\bibitem{nunes2012diagnosis}
V.~D. Nunes, L.~Sawyer, J.~Neilson, G.~Sarri and J.~H. Cross, Diagnosis and
  management of the epilepsies in adults and children: summary of updated nice
  guidance, {\em Bmj} {\bf 344}  (2012).

\bibitem{jenssen2006long}
S.~Jenssen, E.~J. Gracely and M.~R. Sperling, How long do most seizures last? a
  systematic comparison of seizures recorded in the epilepsy monitoring unit,
  {\em Epilepsia} {\bf 47}(9)  (2006)  1499--1503.

\bibitem{goldenberg2010overview}
M.~M. Goldenberg, Overview of drugs used for epilepsy and seizures: etiology,
  diagnosis, and treatment, {\em Pharmacy and Therapeutics} {\bf 35}(7)  (2010)
  p. 392.

\bibitem{world2005atlas}
W.~H. Organization, G.~C. against Epilepsy, P.~for Neurological~Diseases,
  N.~W.~H. Organization), I.~B. for Epilepsy, W.~H. O.~D. of~Mental~Health,
  S.~Abuse, I.~B. of~Epilepsy and I.~L. against Epilepsy, {\em Atlas: epilepsy
  care in the world} (World Health Organization, 2005).

\bibitem{ferri2019ferri}
F.~F. Ferri, {\em Ferri's Clinical Advisor 2020 E-Book: 5 Books in 1} (Elsevier
  Health Sciences, 2019).

\bibitem{berg2008risk}
A.~T. Berg, Risk of recurrence after a first unprovoked seizure, {\em
  Epilepsia} {\bf 49}  (2008)  13--18.

\bibitem{mormann2007seizure}
F.~Mormann, R.~G. Andrzejak, C.~E. Elger and K.~Lehnertz, Seizure prediction:
  the long and winding road, {\em Brain} {\bf 130}(2)  (2007)  314--333.

\bibitem{geut2017detecting}
I.~Geut, S.~Weenink, I.~Knottnerus and M.~J. van Putten, Detecting interictal
  discharges in first seizure patients: ambulatory eeg or eeg after sleep
  deprivation?, {\em Seizure} {\bf 51}  (2017)  52--54.

\bibitem{shah2017optimizing}
V.~Shah, M.~Golmohammadi, S.~Ziyabari, E.~Von~Weltin, I.~Obeid and J.~Picone,
  Optimizing channel selection for seizure detection, {\em 2017 IEEE Signal
  Processing in Medicine and Biology Symposium (SPMB)\/}, , IEEE2017, pp. 1--5.

\bibitem{ayodele2020supervised}
K.~Ayodele, W.~Ikezogwo, M.~Komolafe and P.~Ogunbona, Supervised domain
  generalization for integration of disparate scalp eeg datasets for automatic
  epileptic seizure detection, {\em Computers in Biology and Medicine} {\bf
  120}  (2020) p. 103757.

\bibitem{roy2021evaluation}
S.~Roy, I.~Kiral, M.~Mirmomeni, T.~Mummert, A.~Braz, J.~Tsay, J.~Tang, U.~Asif,
  T.~Schaffter, M.~E. Ahsen {\em et~al.}, Evaluation of artificial intelligence
  systems for assisting neurologists with fast and accurate annotations of
  scalp electroencephalography data, {\em EBioMedicine}   (2021) p. 103275.

\bibitem{furbass2015prospective}
F.~F{\"u}rbass, P.~Ossenblok, M.~Hartmann, H.~Perko, A.~Skupch, G.~Lindinger,
  L.~Elezi, E.~Pataraia, A.~Colon, C.~Baumgartner {\em et~al.}, Prospective
  multi-center study of an automatic online seizure detection system for
  epilepsy monitoring units, {\em Clinical Neurophysiology} {\bf 126}(6)
  (2015)  1124--1131.

\bibitem{mansouri2019online}
A.~Mansouri, S.~P. Singh and K.~Sayood, Online eeg seizure detection and
  localization, {\em Algorithms} {\bf 12}(9)  (2019) p. 176.

\bibitem{gomez2020automatic}
C.~G{\'o}mez, P.~Arbel{\'a}ez, M.~Navarrete, C.~Alvarado-Rojas, M.~Le~Van~Quyen
  and M.~Valderrama, Automatic seizure detection based on imaged-eeg signals
  through fully convolutional networks, {\em Scientific reports} {\bf 10}(1)
  (2020)  1--13.

\bibitem{faust2015wavelet}
O.~Faust, U.~R. Acharya, H.~Adeli and A.~Adeli, Wavelet-based eeg processing
  for computer-aided seizure detection and epilepsy diagnosis, {\em Seizure}
  {\bf 26}  (2015)  56--64.

\bibitem{adeli2003analysis}
H.~Adeli, Z.~Zhou and N.~Dadmehr, Analysis of eeg records in an epileptic
  patient using wavelet transform, {\em Journal of neuroscience methods} {\bf
  123}(1)  (2003)  69--87.

\bibitem{ghosh2007mixed}
S.~Ghosh-Dastidar, H.~Adeli and N.~Dadmehr, Mixed-band wavelet-chaos-neural
  network methodology for epilepsy and epileptic seizure detection, {\em IEEE
  transactions on biomedical engineering} {\bf 54}(9)  (2007)  1545--1551.

\bibitem{savadkoohi2020machine}
M.~Savadkoohi, T.~Oladunni and L.~Thompson, A machine learning approach to
  epileptic seizure prediction using electroencephalogram (eeg) signal, {\em
  Biocybernetics and Biomedical Engineering} {\bf 40}(3)  (2020)  1328--1341.

\bibitem{ansari2019neonatal}
A.~H. Ansari, P.~J. Cherian, A.~Caicedo, G.~Naulaers, M.~De~Vos and
  S.~Van~Huffel, Neonatal seizure detection using deep convolutional neural
  networks, {\em International journal of neural systems} {\bf 29}(04)  (2019)
  p. 1850011.

\bibitem{hu2020scalp}
X.~Hu, S.~Yuan, F.~Xu, Y.~Leng, K.~Yuan and Q.~Yuan, Scalp eeg classification
  using deep bi-lstm network for seizure detection, {\em Computers in Biology
  and Medicine} {\bf 124}  (2020) p. 103919.

\bibitem{bhattacharya2021epileptic}
A.~Bhattacharya, T.~Baweja and S.~Karri, Epileptic seizure prediction using
  deep transformer model, {\em International Journal of Neural Systems}
  (2021) p. 2150058.

\bibitem{raghu2020eeg}
S.~Raghu, N.~Sriraam, Y.~Temel, S.~V. Rao and P.~L. Kubben, Eeg based
  multi-class seizure type classification using convolutional neural network
  and transfer learning, {\em Neural Networks} {\bf 124}  (2020)  202--212.

\bibitem{saab2020weak}
K.~Saab, J.~Dunnmon, C.~R{\'e}, D.~Rubin and C.~Lee-Messer, Weak supervision as
  an efficient approach for automated seizure detection in
  electroencephalography, {\em NPJ digital medicine} {\bf 3}(1)  (2020)  1--12.

\bibitem{emami2019seizure}
A.~Emami, N.~Kunii, T.~Matsuo, T.~Shinozaki, K.~Kawai and H.~Takahashi, Seizure
  detection by convolutional neural network-based analysis of scalp
  electroencephalography plot images, {\em NeuroImage: Clinical} {\bf 22}
  (2019) p. 101684.

\bibitem{nogay2020detection}
H.~S. Nogay and H.~Adeli, Detection of epileptic seizure using pretrained deep
  convolutional neural network and transfer learning, {\em European neurology}
  {\bf 83}(6)  (2020)  602--614.

\bibitem{santaniello2011quickest}
S.~Santaniello, S.~P. Burns, A.~J. Golby, J.~M. Singer, W.~S. Anderson and
  S.~V. Sarma, Quickest detection of drug-resistant seizures: An optimal
  control approach, {\em Epilepsy \& Behavior} {\bf 22}  (2011)  S49--S60.

\bibitem{covert2019temporal}
I.~C. Covert, B.~Krishnan, I.~Najm, J.~Zhan, M.~Shore, J.~Hixson and M.~J. Po,
  Temporal graph convolutional networks for automatic seizure detection, {\em
  Machine Learning for Healthcare Conference\/}, , PMLR2019, pp. 160--180.

\bibitem{roy2019chrononet}
S.~Roy, I.~Kiral-Kornek and S.~Harrer,  Chrononet: a deep recurrent neural
  network for abnormal eeg identification, {\em Conference on Artificial
  Intelligence in Medicine in Europe\/}, , Springer2019, pp. 47--56.

\bibitem{lu2018classification}
Y.~Lu, Y.~Ma, C.~Chen and Y.~Wang, Classification of single-channel eeg signals
  for epileptic seizures detection based on hybrid features, {\em Technology
  and Health Care} {\bf 26}(S1)  (2018)  337--346.

\bibitem{acharya2018deep}
U.~R. Acharya, S.~L. Oh, Y.~Hagiwara, J.~H. Tan and H.~Adeli, Deep
  convolutional neural network for the automated detection and diagnosis of
  seizure using eeg signals, {\em Computers in biology and medicine} {\bf 100}
  (2018)  270--278.

\bibitem{liu2022patient}
G.~Liu, L.~Tian and W.~Zhou, Patient-independent seizure detection based on
  channel-perturbation convolutional neural network and bidirectional long
  short-term memory, {\em International Journal of Neural Systems} {\bf 32}(06)
   (2022) p. 2150051.

\bibitem{shah2021objective}
V.~Shah, M.~Golmohammadi, I.~Obeid and J.~Picone, Objective evaluation metrics
  for automatic classification of eeg events, {\em Biomedical Signal
  Processing}   (2021)  223--255.

\bibitem{reus2022automated}
E.~Reus, G.~Visser, J.~van Dijk and F.~Cox, Automated seizure detection in an
  emu setting: are software packages ready for implementation?, {\em Seizure}
  (2022).

\bibitem{shah2018temple}
V.~Shah, E.~Von~Weltin, S.~Lopez, J.~R. McHugh, L.~Veloso, M.~Golmohammadi,
  I.~Obeid and J.~Picone, The temple university hospital seizure detection
  corpus, {\em Frontiers in neuroinformatics} {\bf 12}  (2018) p.~83.

\bibitem{shoeb2004patient}
A.~Shoeb, H.~Edwards, J.~Connolly, B.~Bourgeois, S.~T. Treves and J.~Guttag,
  Patient-specific seizure onset detection, {\em Epilepsy \& Behavior} {\bf
  5}(4)  (2004)  483--498.

\bibitem{stevenson2019dataset}
N.~Stevenson, K.~Tapani, L.~Lauronen and S.~Vanhatalo, A dataset of neonatal
  eeg recordings with seizure annotations, {\em Scientific data} {\bf 6}(1)
  (2019)  1--8.

\bibitem{burrello2019hyperdimensional}
A.~Burrello, K.~Schindler, L.~Benini and A.~Rahimi, Hyperdimensional computing
  with local binary patterns: one-shot learning of seizure onset and
  identification of ictogenic brain regions using short-time ieeg recordings,
  {\em IEEE Transactions on Biomedical Engineering} {\bf 67}(2)  (2019)
  601--613.

\bibitem{wagenaar2013multimodal}
J.~B. Wagenaar, B.~H. Brinkmann, Z.~Ives, G.~A. Worrell and B.~Litt,  A
  multimodal platform for cloud-based collaborative research, {\em 2013 6th
  international IEEE/EMBS conference on neural engineering (NER)\/}, ,
  IEEE2013, pp. 1386--1389.

\bibitem{li2021neural}
A.~Li, C.~Huynh, Z.~Fitzgerald, I.~Cajigas, D.~Brusko, J.~Jagid, A.~O. Claudio,
  A.~M. Kanner, J.~Hopp, S.~Chen {\em et~al.}, Neural fragility as an eeg
  marker of the seizure onset zone, {\em Nature neuroscience} {\bf 24}(10)
  (2021)  1465--1474.

\bibitem{thomas2021automated}
J.~Thomas, P.~Thangavel, W.~Y. Peh, J.~Jing, R.~Yuvaraj, S.~S. Cash,
  R.~Chaudhari, S.~Karia, R.~Rathakrishnan, V.~Saini {\em et~al.}, Automated
  adult epilepsy diagnostic tool based on interictal scalp electroencephalogram
  characteristics: A six-center study, {\em International Journal of Neural
  Systems}   (2021) p. 2050074.

\bibitem{thangavel2021time}
P.~Thangavel, J.~Thomas, W.~Y. Peh, J.~Jing, R.~Yuvaraj, S.~S. Cash,
  R.~Chaudhari, S.~Karia, R.~Rathakrishnan, V.~Saini {\em et~al.},
  Time--frequency decomposition of scalp electroencephalograms improves deep
  learning-based epilepsy diagnosis, {\em International Journal of Neural
  Systems}   (2021) p. 2150032.

\bibitem{peh2021multi}
W.~Y. Peh, J.~Thomas, E.~Bagheri, R.~Chaudhari, S.~Karia, R.~Rathakrishnan,
  V.~Saini, N.~Shah, R.~Srivastava, Y.-L. Tan {\em et~al.}, Multi-center
  validation study of automated classification of pathological slowing in adult
  scalp electroencephalograms via frequency features, {\em International
  Journal of Neural Systems}   (2021) p. 2150016.

\bibitem{joo2020being}
T.~Joo, U.~Chung and M.-G. Seo,  Being bayesian about categorical probability,
  {\em International Conference on Machine Learning\/}, , PMLR2020, pp.
  4950--4961.

\bibitem{peh2022transformer}
W.~Y. Peh, Y.~Yao and J.~Dauwels,  Transformer convolutional neural networks
  for automated artifact detection in scalp eeg, {\em 2022 44th Annual
  International Conference of the IEEE Engineering in Medicine \& Biology
  Society (EMBC)\/}, , IEEE2022, pp. 3599--3602.

\bibitem{guo2017calibration}
C.~Guo, G.~Pleiss, Y.~Sun and K.~Q. Weinberger,  On calibration of modern
  neural networks, {\em International Conference on Machine Learning\/}, ,
  PMLR2017, pp. 1321--1330.

\bibitem{afra2008duration}
P.~Afra, C.~C. Jouny and G.~K. Bergey, Duration of complex partial seizures: an
  intracranial eeg study, {\em Epilepsia} {\bf 49}(4)  (2008)  677--684.

\bibitem{cook2013prediction}
M.~J. Cook, T.~J. O'Brien, S.~F. Berkovic, M.~Murphy, A.~Morokoff, G.~Fabinyi,
  W.~D'Souza, R.~Yerra, J.~Archer, L.~Litewka {\em et~al.}, Prediction of
  seizure likelihood with a long-term, implanted seizure advisory system in
  patients with drug-resistant epilepsy: a first-in-man study, {\em The Lancet
  Neurology} {\bf 12}(6)  (2013)  563--571.

\bibitem{o2020neonatal}
A.~O’Shea, G.~Lightbody, G.~Boylan and A.~Temko, Neonatal seizure detection
  from raw multi-channel eeg using a fully convolutional architecture, {\em
  Neural Networks} {\bf 123}  (2020)  12--25.

\bibitem{brinkmann2016crowdsourcing}
B.~H. Brinkmann, J.~Wagenaar, D.~Abbot, P.~Adkins, S.~C. Bosshard, M.~Chen,
  Q.~M. Tieng, J.~He, F.~Mu{\~n}oz-Almaraz, P.~Botella-Rocamora {\em et~al.},
  Crowdsourcing reproducible seizure forecasting in human and canine epilepsy,
  {\em Brain} {\bf 139}(6)  (2016)  1713--1722.

\bibitem{sierra2015seizure}
A.~Sierra-Marcos, M.~L. Scheuer and A.~O. Rossetti, Seizure detection with
  automated eeg analysis: a validation study focusing on periodic patterns,
  {\em Clinical neurophysiology} {\bf 126}(3)  (2015)  456--462.

\bibitem{rommens2018improving}
N.~Rommens, E.~Geertsema, L.~J. Holleboom, F.~Cox and G.~Visser, Improving
  staff response to seizures on the epilepsy monitoring unit with online eeg
  seizure detection algorithms, {\em Epilepsy \& Behavior} {\bf 84}  (2018)
  99--104.

\bibitem{koren2021systematic}
J.~Koren, S.~Hafner, M.~Feigl and C.~Baumgartner, Systematic analysis and
  comparison of commercial seizure-detection software, {\em Epilepsia} {\bf
  62}(2)  (2021)  426--438.

\bibitem{asif2020seizurenet}
U.~Asif, S.~Roy, J.~Tang and S.~Harrer, Seizurenet: Multi-spectral deep feature
  learning for seizure type classification, {\em Machine Learning in Clinical
  Neuroimaging and Radiogenomics in Neuro-oncology}   (2020)  77--87.

\bibitem{yuan2018multi}
Y.~Yuan, G.~Xun, K.~Jia and A.~Zhang, A multi-view deep learning framework for
  eeg seizure detection, {\em IEEE journal of biomedical and health
  informatics} {\bf 23}(1)  (2018)  83--94.

\bibitem{zhou2018epileptic}
M.~Zhou, C.~Tian, R.~Cao, B.~Wang, Y.~Niu, T.~Hu, H.~Guo and J.~Xiang,
  Epileptic seizure detection based on eeg signals and cnn, {\em Frontiers in
  neuroinformatics} {\bf 12}  (2018) p.~95.

\bibitem{pereira2020fema}
D.~R. Pereira, M.~A. Piteri, A.~N. Souza, J.~P. Papa and H.~Adeli, Fema: a
  finite element machine for fast learning, {\em Neural Computing and
  Applications} {\bf 32}(10)  (2020)  6393--6404.

\bibitem{alam2020dynamic}
K.~M.~R. Alam, N.~Siddique and H.~Adeli, A dynamic ensemble learning algorithm
  for neural networks, {\em Neural Computing and Applications} {\bf 32}(12)
  (2020)  8675--8690.

\end{thebibliography}

\end{multicols}
\end{document}